\documentclass[twocolumn,english,aps,prb,floatfix,preprintnumbers,showpacs,amsfonts,amssymb,superscriptaddress]{revtex4}
\usepackage{ae,aecompl}
\usepackage[T1]{fontenc}
\usepackage[latin1]{inputenc}
\usepackage{color}
\usepackage{amsmath}
\usepackage{amssymb}
\usepackage{graphicx}

\makeatletter

\providecommand{\tabularnewline}{\\}

\@ifundefined{textcolor}{}
{%
 \definecolor{BLACK}{gray}{0}
 \definecolor{WHITE}{gray}{1}
 \definecolor{RED}{rgb}{1,0,0}
 \definecolor{GREEN}{rgb}{0,1,0}
 \definecolor{BLUE}{rgb}{0,0,1}
 \definecolor{CYAN}{cmyk}{1,0,0,0}
 \definecolor{MAGENTA}{cmyk}{0,1,0,0}
 \definecolor{YELLOW}{cmyk}{0,0,1,0}
}

\usepackage{amscd}
\usepackage{bm}
\usepackage{graphics,psfrag}
\usepackage{graphicx,psfrag}

\makeatother

\usepackage{babel}
\begin{document}

\title{Entanglement Entropy of the Low-Lying Excited States and Critical
Properties of an Exactly Solvable Two-Leg Spin Ladder with Three-Spin
Interactions}

\author{D.~Eloy}

\affiliation{Instituto de F\'{\i}sica, Universidade Federal de Uberlândia, Caixa
Postal 593, 38400-902 Uberlândia, MG, Brazil }

\author{J.~C.~Xavier}

\affiliation{Instituto de F\'{\i}sica, Universidade Federal de Uberlândia, Caixa
Postal 593, 38400-902 Uberlândia, MG, Brazil }

\date{\today{}}
\begin{abstract}
In this work, we investigate an exactly solvable two-leg spin ladder
with three-spin interactions. We obtain analytically the finite-size
corrections of the low-lying energies and determine the central charge
as well as the scaling dimensions. The model considered in this work 
is in the same universality
class of critical behavior of the $XX$ chain with central charge
$c=1$. By using the correlation matrix method, we also study  the
finite-size corrections of the Rényi entropy of the ground state and
of the excited states. Our results are in agreement with the predictions
of the conformal field theory. 
\end{abstract}


\maketitle

\section{Introduction}

Perhaps one of the most popular and precise ways to determine the critical
behavior of one-dimensional quantum systems is through the finite-size
corrections of the energy. The machinery of  conformal field theory
(CFT) establishes that the ground state energy of a one-dimensional
critical system of size $L$, under periodic boundary condition (PBC),
behaves asymptotically as\citealp{anomaly1,anomaly2,critical3}

\begin{equation}
\frac{E_{0}}{L}=e_{\infty}-\frac{v_{s}\pi c}{6L^{2}}+o(L^{-2}),\label{eq:1}
\end{equation}
where $v_{s}$ is the sound velocity, $e_{\infty}$ is the bulk ground
state energy per site, and $c$ is the central charge.

The mass gap amplitudes of the finite-size corrections of the higher
energy states, are related with the scaling dimensions $d^{\beta}$.
There are, for each primary operator $O_{\beta}$ ($\beta=1,2,...)$
in the CFT, a tower of states $E_{j,j'}^{\beta}(L)$ in the spectrum
of the Hamiltonian with asymptotic behavior \citealp{dimensions,critical3}

\begin{equation}
E_{j,j'}^{\beta}(L)-E_{0}(L)=\frac{2\pi v_{s}}{L}(d^{\beta}+j+j')+o(L^{-1}),\label{eq:2}
\end{equation}
 where $j,j'=0,1,2..$. . The above relations were used systematically
to determine the universality class of critical behavior of several
models with great success. The success of these relations resides
in the fact that it is possible to obtain the critical exponents,
which are associated with physical quantities in the thermodynamic
limit, from \emph{finite} systems.

Recently, a great deal of excitement has been generated due to the
fact that the universality classes of critical behavior of one-dimensional
systems can also be inferred from the finite-size corrections of the
entanglement entropy of the ground state as well as of the excited
states. \citep{cvidal,revfazio,prlkorepin,cardyentan,chicoentropy}
Below, we briefly report some important results about these corrections
that we will explore in this work.

Consider a quantum chain with $L$ sites, described by a pure state
whose density operator is $\rho$. Let us consider that the system
is composed by the subsystems ${\cal A}$ with $\ell$ sites ($\ell=1,\ldots,L-1$)
and ${\cal B}$ with $L-\ell$ sites. The Rényi entropy is defined
as 

\begin{equation}
S_{\alpha}(L,\ell)=\frac{1}{1-\alpha}\ln Tr(\rho_{{\cal {A}}}^{\alpha}),\label{eq:renyientropy}
\end{equation}
where $\rho_{{\cal {A}}}=\mbox{Tr}_{{\cal {B}}}\rho$ is the reduced
density matrix of the subsystem ${\cal {A}}$. The von Neumann entropy,
also known as entanglement entropy, is the particular case $\alpha=1$.
In the scaling regime $1<<\ell<<L$, it is expected that for the critical
one-dimensional systems with PBC, the Rényi entropy of the ground
state behaves as

\begin{equation}
S_{\alpha}(L,\ell)=S_{\alpha}^{CFT}(L,\ell)+S_{\alpha}^{osc}(L,\ell).\label{eq:entropyb}
\end{equation}
The first term in this equation, is the CFT prediction and is given
by \citealp{cold,cardyentan,entroreviewcalabrese,affleckboundary} 

\begin{equation}
S_{\alpha}^{CFT}=\frac{c}{6}\left(1+\frac{1}{\alpha}\right)\ln\left[\frac{L}{\pi}\sin\left(\frac{\pi\ell}{L}\right)\right]+c_{\alpha},\label{eq:entropyCFT}
\end{equation}
where $c_{\alpha}$ is a non-universal constant. The second term is
given by\citealp{entropyosc,xxPBCh,calabreseOBC} 

\begin{equation}
S_{\alpha}^{osc}=\frac{\left[a_{1}\delta_{1,\alpha}+g_{\alpha}(1-\delta_{1,\alpha})\cos(\kappa\ell+\phi)\right]}{L^{p_{\alpha}}}\left|\sin\left(\pi\frac{\ell}{L}\right)\right|^{-p_{\alpha}},\label{eq:entropyUnusual}
\end{equation}
where $p_{\alpha}$ is a new critical exponent, and $a_{1}$ and $g_{\alpha}$
are  non-universal constants. The wave vector $\kappa$ and the phase
$\phi$ depend on the model. For instance, for the Ising model $\kappa=0=\phi$,
and for the spin-$s$ $XXZ$ chains at zero magnetic field $\kappa=\pi$
and $\phi=0$. For the systems with PBC, it is not expected oscillations
in the von Neumann entropy $S_{1}.$\citealp{entropyosc,XavierAlca2012}
The origin of the exponent $p_{\alpha}$ in the above equation are
the conical space-time singularities produced in the conformal mapping
used to describe the reduced density matrix $\rho_{{\cal {A}}}=\mbox{Tr}_{{\cal {B}}}\rho$
in the CFT.\citealp{cardyosc} This exponent is related to the scaling
dimension $x^{con}$ of an operator (unknown) of the underlying CFT
as follows $p_{\alpha}=2x^{con}/\alpha$.\citealp{cardyosc} There are evidences
that the exponent $x^{con}=x_{\epsilon}$ for $\alpha>1$, where $x_{\epsilon}$
is the scaling dimension of the energy operator, and that $p_{1}=\nu=2$
for all models.\citealp{XavierAlca2012} On the other hand, the origin
of the oscillating factor {[}$\cos(\kappa\ell+\phi)${]} is not yet
completely understood, however it has been observed in systems with
anti-ferromagnetic tendencies. 

The CFT prediction for the finite-size corrections of the Rényi entropy
above were verified by several authors. 
Both the central charge\citealp{entropyaffleckosc,peterentropy,xavierentanglement,entropyosc,xxPBCh,calabreseOBC,calabreserandom}
and the exponent $p_{\alpha}$ \citealp{entropyosc,calabreseOBC,XavierAlca2012,xavieralcarazosc,taddiaosc,unsu2012}
were obtained with amazing precision.

The Rényi entropies of the excited states also present very interesting
universal behavior.\citealp{chicoentropy,chicoentropyext2,CFThomog}
The $\alpha-$Rényi entropy $S_{\alpha}^{\Upsilon}$, associated with
an excitation of a primary operator $\Upsilon$ with conformal weight
$h$, is related with the $2\alpha-$point correlator of the operator
$\Upsilon$ and its conjugate by\citealp{chicoentropy,chicoentropyext2}

\[
\ln F_{\Upsilon}^{(\alpha)}=(1-\alpha)[S_{\alpha}^{\Upsilon}-S_{\alpha}^{gs}],
\]
where $S_{\alpha}^{gs}$ is the $\alpha-$Rényi entropy of the ground
state and
\[
F_{\Upsilon}^{(\alpha)}=\alpha^{-2\alpha(h+\bar{h})}\frac{\left\langle \prod_{j=0}^{\alpha-1}\Upsilon[2\pi j/\alpha]\Upsilon^{\dagger}[2\pi(j+\ell/L)/\alpha]\right\rangle }{\left\langle \Upsilon[0]\Upsilon^{\dagger}[2\pi\ell/L]\right\rangle ^{\alpha}}.
\]
 In particular, $F_{\Upsilon}^{(\alpha)}=1$ for the operators associated
with compact excitations,\citealp{chicoentropy,chicoentropyext2}
and the Rényi entropy of these excitations are the same of the ground
state, i. e. , $S_{\alpha}^{\Upsilon}=S_{\alpha}^{gs}$. On the other
hand, for non-compact excitations the universal function $F_{\Upsilon}^{(\alpha)}$
depends on the particular primary operator $\Upsilon$. For instance,
for the primary operator $\Upsilon=i\partial\phi$ of a free boson
theory the universal function $F_{i\partial\phi}^{(2)}$ was calculated
analytically and is given by\citealp{chicoentropy,chicoentropyext2}
\begin{equation}
F_{i\partial\phi}^{(2)}=1-2s^{2}+3s^{4}-2s^{6}+8s^{8}\label{eq:F2}
\end{equation}
where $s=\sin(\pi\ell/2L)$. 

In this work, we intend to explore the above relations in order to
investigate a two-leg spin ladder model, which is exactly soluble. The paper
is organized as follows. In  Sec. II, we present the model and
its ground state phase diagram. In  Sec. III, we determine the
Rényi entropy of the model by using the correlation matrix method.
Finally, in Sec. IV we summarize our results.

\section{Phase Diagram and critical properties of a two-leg spin model}

\begin{figure}
\begin{centering}
\psfrag{ell}{$\ell$}
\includegraphics[scale=0.08]{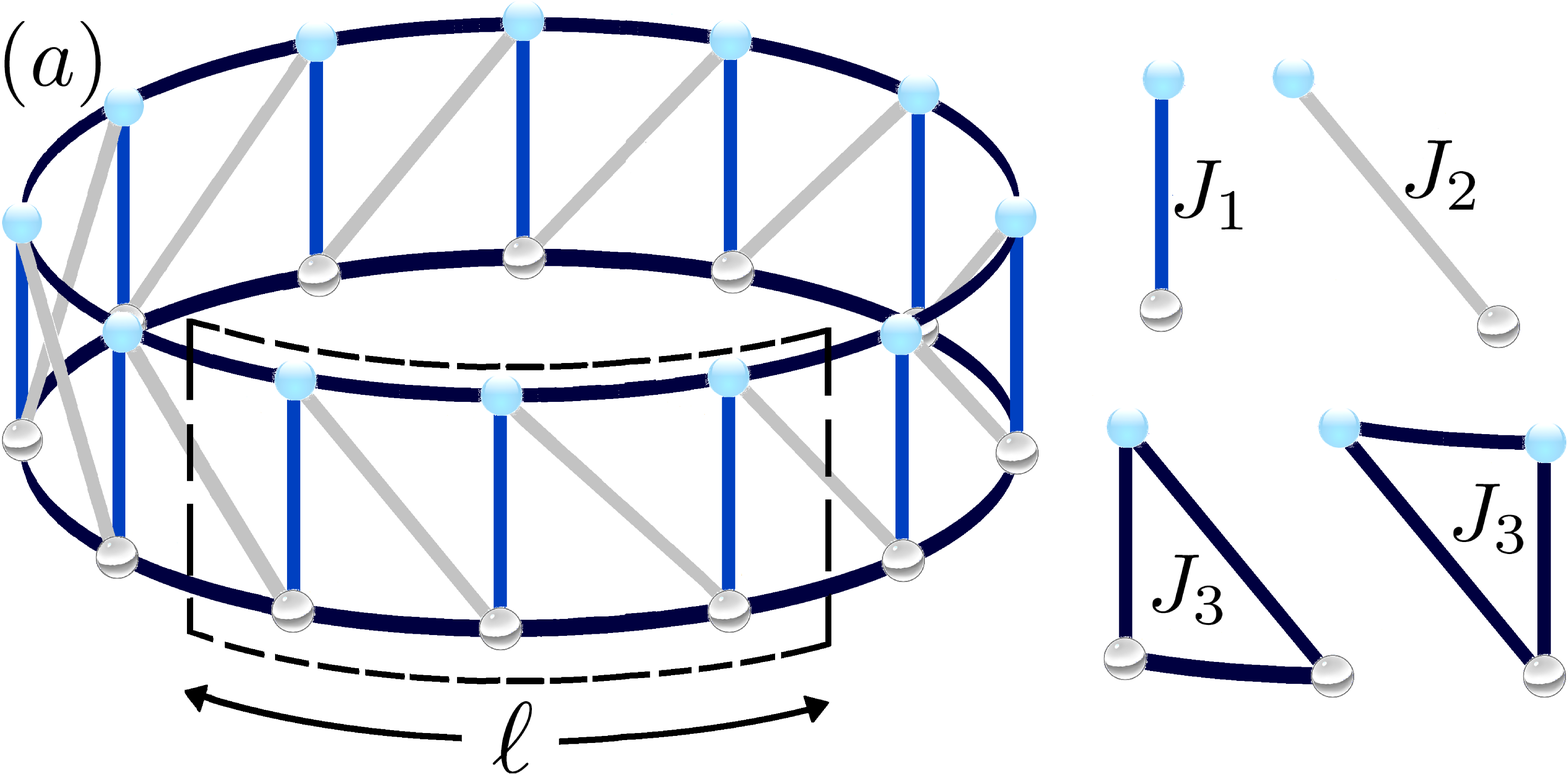}
\vspace*{0.5cm}
\end{centering}
\begin{centering}
\includegraphics[scale=0.08]{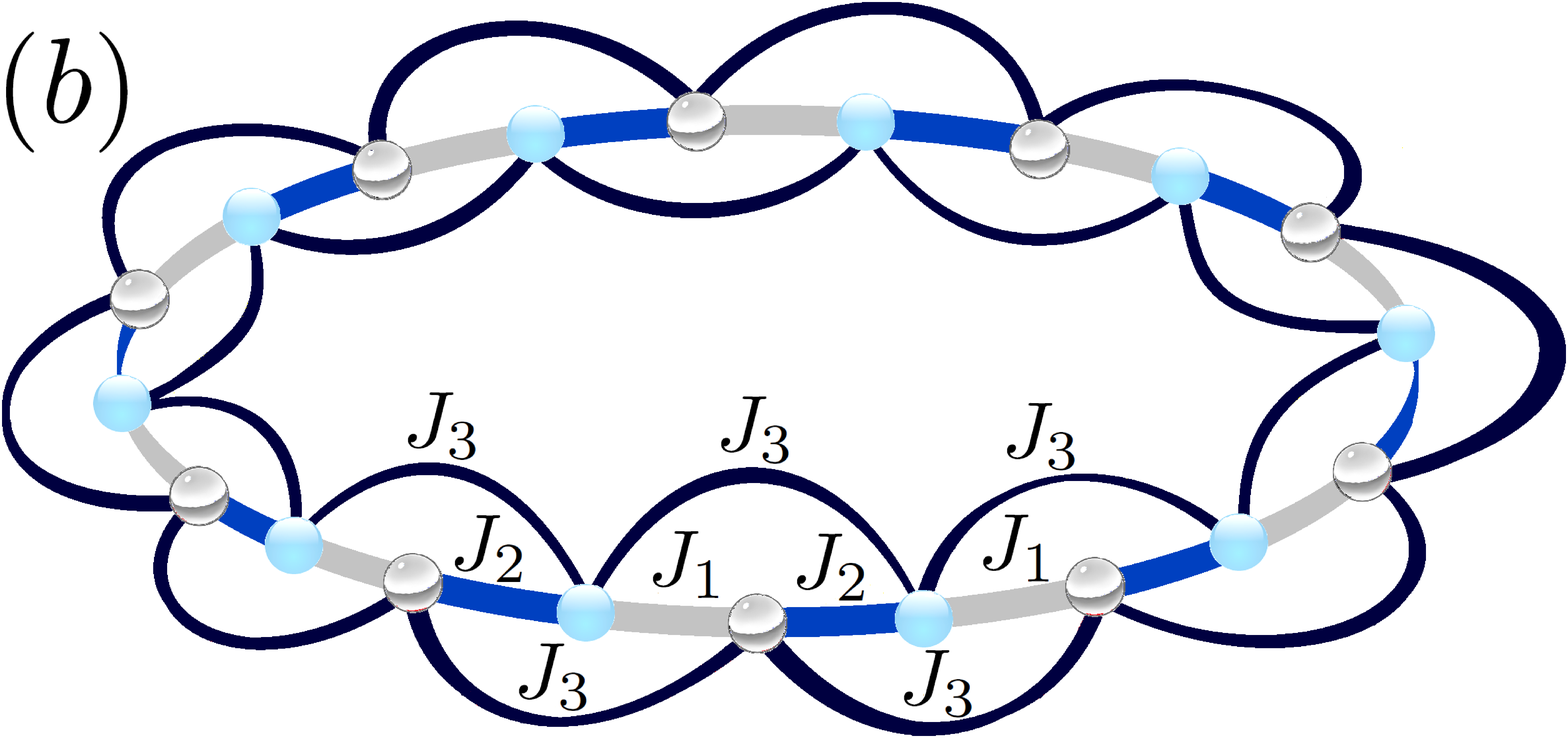}
\end{centering}
\caption{\label{fig1} (Color online). (a) Schematic representation of the
two-leg spin ladder. $J_{1},$ $J_{2}$ and $J_{3}$ are the couplings
between the spins along the rung, the diagonal and the plaquettes, respectively.
We considered a bipartite system as reported in the figure. The system
of $L$ rungs is divided into two subsystems of sizes $\ell$ and
$(L-\ell)$. (b) The spinless Hamiltonian (\ref{eq:Hfermions}). The
lines represent the hopping terms.}
\end{figure}

We consider the following Hamiltonian defined on a two-leg geometry

\[
H=2\sum_{\beta=x,y}\sum_{\lambda=1}^{2}\sum_{j=1}^{L}(J_{\lambda}s_{\lambda,j}^{\beta}s_{\lambda+1,j+\lambda-1}^{\beta}+
\]
\begin{equation}
J_{3}s_{\lambda,j}^{\beta}s_{\lambda+1,j+\lambda-1}^{z}s_{\lambda,j+1}^{\beta})-h\sum_{\lambda=1}^{2}\sum_{j=1}^{L}s_{\lambda,j}^{z},\label{Eq:Hamiltonian}
\end{equation}
where $J_{j}(j=1,2,3)$ are coupling constants {[}see Fig. \ref{fig1}(a){]},
$s_{\lambda,j}^{\beta}(\beta=x,y,z)$ are the spin-$\frac{1}{2}$
operators at leg $\lambda=1,2$ and rung $j$, $h$ is the magnetic
field, and $L$ is considered even. We investigate the above model
with PBC, i. e., $s_{\lambda,j+L}^{\beta}=s_{\lambda,j}^{\beta}$
and $s_{\lambda+2,j}^{\beta}=s_{\lambda,j}^{\beta}$. Note that this
model has three-spin interactions. Models with multi-spin interactions
like $s_{j}^{\beta}\prod_{l=j+1}^{m}s_{l}^{z}$$s_{m+1}^{\beta'}$
(with $\beta,\beta'=x,y$) can be mapped into a fermionic quadratic
form, as first noted by Suzuki. \citep{suzukixy} In the particular
case of three-spin interactions, several variants of the Hamiltonian
(\ref{Eq:Hamiltonian}) were considered in the literature\citep{3body2003,3body2004,3body2009,3body2011,3bodyDerzhko2009,3body2012,3bodyfagotti}
(see also Ref. \onlinecite{3bodybethe1990b}). It is quite interesting
to mention that spin-$\frac{1}{2}$ Hamiltonians with three-spin interactions
can be generated using optical lattices.\citep{3bodyoptical}

The authors of Ref. \onlinecite{3bodyDerzhko2009} also investigated
the same model considered in this work. However, these authors focused
in the effect of the lattice distortion in the infinity system. Here,
our emphasis is in the \emph{finite-size corrections} (i) of the entanglement
entropy and (ii) of the low-lying energy states in the critical regions.
As we already mentioned, these corrections are related with the central
charge and the scaling dimensions.

\begin{figure}
\begin{centering}
\includegraphics[scale=0.12]{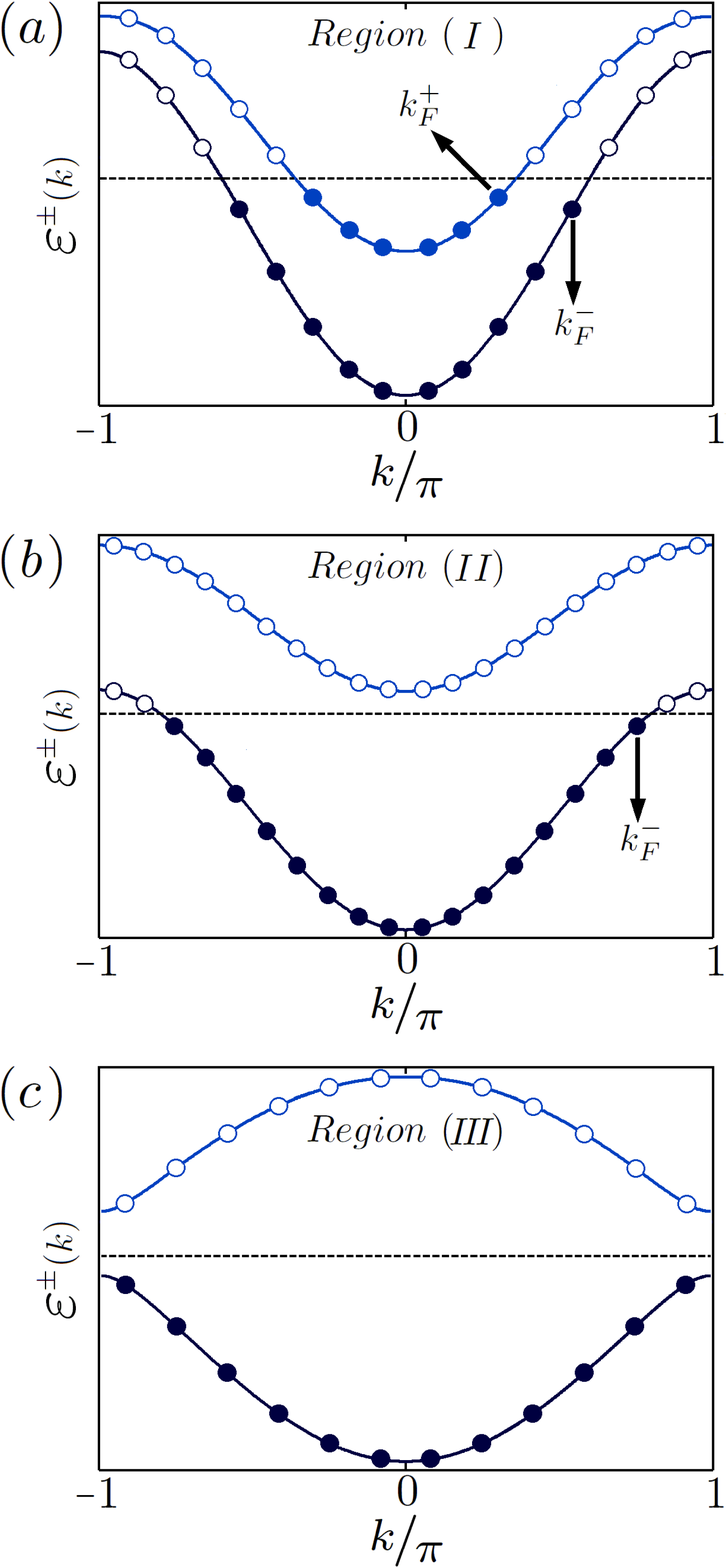}
\par\end{centering}

\caption{\label{fig2} (Color online). Figures (a), (b) and (c) display the
three typical profiles of the band dispersion, for $h=0$, that appear
in the regions I, II and III, respectively {[}see Fig. (\ref{fig3}){]}.
$k_{F}^{\pm}$ are the Fermi momenta in the branches $\sigma=\pm$
(see text). }
\end{figure}

For the sake of clarity, we briefly describe the procedure to diagonalize
the Hamiltonian (\ref{Eq:Hamiltonian}) exactly. First, we use a Jordan-Wigner
transformation {[}$c_{m}=\left(s_{m}^{x}-is_{m}^{y}\right)\prod_{j<m}(-2s_{j}^{z})${]}
in order to map the Hamiltonian (\ref{Eq:Hamiltonian}) in a spinless
fermion chain with nearest-neighbor and next-neighbor hoppings {[}see
Fig. \ref{fig1}(b){]}. By using this transformation we obtain 
\[
H=\sum_{i=1}^{L-1}\left(J_{1}c_{_{2i-1}}^{\dagger}c_{_{2i}}+J_{2}c_{_{2i}}^{\dagger}c_{_{2i+1}}\right)-\sum_{j=1}^{2L-2}\frac{J_{3}}{2}c_{_{i}}^{\dagger}c_{_{i+2}}+
\]
\[
J_{1}c_{_{2L-1}}^{\dagger}c_{_{2L}}-e^{i\pi\hat{\mathcal{N}}_{F}}\left[J_{2}c_{_{2L}}^{\dagger}c_{_{1}}-\frac{J_{3}}{2}\left(c_{_{2L-1}}^{\dagger}c_{_{1}}+c_{_{2L}}^{\dagger}c_{_{2}}\right)\right]+h.c.
\]

\begin{equation}
-h\hat{\mathcal{N}}_{F}+hL,\label{eq:Hfermions}
\end{equation}
where $\hat{\mathcal{N}}_{F}=\sum_{i=1}^{2L}c_{i}^{\dagger}c_{i}$
is the particle number operator. Note that the original two-leg ladder system of size $L$ 
is mapped into a chain with $2L$ sites.

Finally, by using the following Fourier
transformations

\begin{equation}
c_{_{2j-1}}=\sum_{k}\frac{e^{ikj}}{\sqrt{L}}a_{k}\textrm{ and }c_{_{2j}}=\sum_{k}\frac{e^{ikj}}{\sqrt{L}}b_{k},\label{eq:Fourier}
\end{equation}
we can express the Hamiltonian (\ref{eq:Hfermions}) in the following
diagonal form

\begin{equation}
H=\sum_{\sigma=\pm}\sum_{\{k\}}\varepsilon^{\sigma}(k)d_{k}^{\sigma}{}^{\dagger}d_{k}^{\sigma}+hL,\label{eq:Hdiag}
\end{equation}
where $d_{k}^{\pm}=\frac{1}{\sqrt{2}}\left(a_{k}\pm\frac{J_{1}+J_{2}\exp(-ik)}{\sqrt{J_{1}^{2}+J_{2}^{2}+2J_{1}J_{2}\textrm{cos}k}}b_{k}\right)$,
$k=k(j)=\frac{2j\pi+\phi}{L}$ with $\phi=0$ ($\pi$) if $\mathcal{N}_{F}$
is odd (even) and $j=0,\pm1,...,\pm(L-2)/2,-L/2$. The band dispersion
is separated in two branches $(\sigma=\pm)$ given by
\begin{equation}
\varepsilon^{\sigma}(k)=-h-J_{3}\textrm{cos}k+\sigma\sqrt{J_{1}^{2}+J_{2}^{2}+2J_{1}J_{2}\textrm{cos}k}.\label{eq:bands}
\end{equation}

\begin{figure}
\begin{centering}
\includegraphics[scale=0.22]{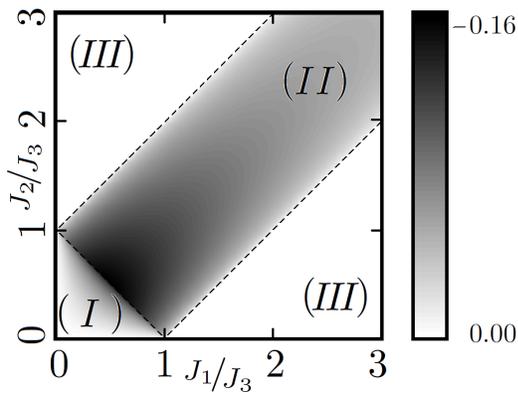}
\par\end{centering}

\caption{\label{fig3} (Color online). The ground state phase diagram for $h=0$.
The regions I, II are gapless, whereas the region III is gapped. The
right panel indicates the intensity of the magnetization per site
$m$. }
\end{figure}

Let us focus in the case of a zero magnetic field. After an exhaustive
investigation of the band dispersion above with $h=0$, we note that
depending on the coupling constants $J_{1}$, $J_{2}$, and $J_{3}$,
the profiles of the band dispersion present three distinct behaviors.
The main characteristics of these profiles are depicted in Figs. \ref{fig2}(a)-(c).
We label the coupling regions by regions I, II and III according to
the Fermi momenta in the thermodynamic limit. In the region I (II)
we have four (two) Fermi momenta, while in the region III the Fermi
momenta always appear at $k_{F}^{-}=\pm(\pi-\pi/L)$. We can observe
from Fig. \ref{fig2} that the regions I and II are gapless, whereas
the region III is gapped. In the latter case, the lowest excitation
energy appears in the sector $s_{total}^{z}=-1$. The spin gap in
this region is given by $\Delta=|J_{1}-J_{2}|-J_{3}$. \citep{comment2012}
Fig. \ref{fig3} displays these regions in parameter space
$J_{2}/J_{3}$ vs $J_{1}/J_{2}$.

It is interesting also to note that depending on the coupling constants 
the magnetization of the ground state may not be zero in regions I and II,
differently from the region III where the magnetization is always zero if $h=0$.
The magnetization per site is given by $\frac{m}{2L}=\frac{1}{2L}\sum_{j}<gs|\sum s_{j}^{z}|gs>=\frac{k_{F}^{+}+k_{F}^{-}}{2\pi}-1/2$,
where the Fermi momenta $k_{F}^{\pm}$ are determined solving the
equations $\epsilon^{\pm}(k_{F}^{\pm})=0$, in the thermodynamic limit,
and are given by 
\[
\ensuremath{k_{F}^{\pm}=}\arccos\Bigg(\frac{J_{1}J_{2}-hJ_{3}}{J_{3}^{2}}\pm
\]

\begin{equation}
\frac{\sqrt{J_{1}^{2}J_{2}^{2}+J_{1}^{2}J_{3}^{2}+J_{2}^{2}J_{3}^{2}-2J_{1}J_{2}J_{3}h}}{J_{3}^{2}}\Bigg).\label{eq:KFs}
\end{equation}
From the above equations we can see that for $h=0$ the magnetization
is not zero, in general. In Fig. \ref{fig3}, we also report the values
of the intensity of magnetization $m$ in the parameter space. 

Let us now investigate the critical regions I and II in more detail.\textcolor{red}{{}
}The ground state energy of the Hamiltonian (\ref{eq:Hdiag})
is obtained adding particles in the energy levels below zero, i. e.,
\begin{equation}
E_{0}-hL=\sum_{|k|\le k_{F}^{-}}\varepsilon^{-}(k)+\sum_{|k|\le k_{F}^{+}}\varepsilon^{+}(k),\label{eq:Egs}
\end{equation}
where the last sum in this equation contributes only for the region
I. We consider $\mathcal{N}_{F}=\mathcal{N}_{F}^{-}+\mathcal{N}_{F}^{+}$
even and $k_{F}^{\pm}=(\mathcal{N}_{F}^{\pm}-1)\pi/L$ are the Fermi
momenta, where $\mathcal{N}_{F}^{\alpha}$ is the particle number
in the branch $\sigma$. 

In order to determine the central charge $c$, we need to evaluate
the finite-size corrections of the ground state energy {[}Eq. (\ref{eq:Egs}){]}.
For a fixed value of the density $\rho=\frac{\mathcal{N}_{F}}{2L}$
the leading finite-size correction can be obtained by using the Euler-Maclaurin
formula,\citep{bookAbramo} which gives 

\begin{equation}
\frac{E_{0}}{L}=e_{\infty}-\frac{\pi\left(\upsilon_{F}^{+}+\upsilon_{F}^{-}\right)}{6L^{2}}+
{o}\left(L^{-2}\right),\label{eq:Egsfinite}
\end{equation}
where $\footnotesize\upsilon_{F}^{\pm}=\frac{d\varepsilon^{\pm}}{dk}|_{k_{F}^{\pm}}=(J_{3}\mp\frac{J_{1}J_{2}}{\sqrt{J_{1}^{2}+J_{2}^{2}+2J_{1}J_{2}\textrm{cos}(2\pi\rho^{\pm})}})\sin(2\pi\rho^{\pm})$
are the sound velocities associated with each branch, 
$\rho^{\pm}=\frac{\mathcal{N}_{F}^{\pm}}{2L}$,
and $e_{\infty}=e_{\infty}^{+}+e_{\infty}^{-}$ is the bulk energy
with $e_{\infty}^{\pm}$ given by

\[
e_{\infty}^{\pm}=h\left(\frac{1}{2}-2\rho^{\pm}\right)-\frac{J_{3}}{\pi}\textrm{sin}(2\pi\rho^{\pm})\pm
\]

\begin{equation}
\int_{0}^{2\pi\rho^{\pm}}\frac{dk}{\pi}\sqrt{J_{1}^{2}+J_{2}^{2}+2J_{1}J_{2}\textrm{cos}k}.\label{eq:einftotal}
\end{equation}
If we compare Eqs. (\ref{eq:1}) and (\ref{eq:Egsfinite}), we identify
the central charge as $c=1$ for each gapless mode. This value of $c$ implies that the
universality class of critical behavior of the Hamiltonian (\ref{Eq:Hamiltonian})
is the same of the well known $XX$ chain. 
Note that 
the critical behavior is described by two \emph{non-interacting}
conformal theories associated with the two branches $\sigma=\pm$.

The excited states are obtained adding particles (holes) above (below)
the Fermi level. Let us consider the following excitations: (i) we
can add (remove) $I=I_{+}+I_{-}$ particles above (below) the Fermi
level, where $I_{\sigma}$ $(\sigma=\pm)$ is the number of particles
added (removed) in the branch $\sigma$; and/or (ii) we can remove
$Q_{\sigma}$  particles in highest occupied level of the left/right Fermi
point of the branch $\sigma$ and add these particles 
in the lowest free levels of the right/left Fermi point of the same  branch  (see Table I). 
The finite-size corrections of these states can also be obtained 
by using the Euler-Maclaurin formula, as done for the ground state energy. 
The result obtained is the following

\begin{equation}
E^{\{I_{\sigma}\},\{Q_{\sigma}\}}-E_{0}=
\sum_{\sigma=\pm}\frac{2\pi\upsilon_{F}^{\sigma}}{L}
 \left[\frac{1}{4} I_{\sigma}^2+\left(Q_{\sigma}+\delta_\sigma/2 \right)^2\right]+o(L^{-2}).\label{eq:exctfinite}
\end{equation}
where  $\delta_{\sigma}=\mod\left[\mod(I_{\sigma},2)+\mod(I,2),2\right]$ in the region I and
$\delta_{\sigma}=0$ in the region II (note that in this region
$\upsilon_{F}^{+}=0$). 
We observe, again, from the above equation 
that the Hamiltonian (\ref{Eq:Hamiltonian}) is described by two \emph{non-interacting}
conformal theories. Similar expressions also have been
found in models with more than one gapless mode. \citep{frahmkorepin,korepinNcompJA,woynarovichHubbard,EsslerHubbard,marcioSp2n}
Comparing  Eqs. (\ref{eq:2}) and (\ref{eq:exctfinite}) we identify
the scaling dimensions as 
\begin{equation}
d_{I_{\sigma},Q_{\sigma}}^{\sigma} =\frac{1}{4}I_{\sigma}^{2}+\left(Q_{\sigma}+\delta_\sigma/2 \right)^2.
\label{eq:scaldim}
\end{equation}
The scaling dimensions above have a similar structure of the one of the
Gaussian model.\citealp{critical3}
Note that $d^{\sigma}_{0,1}=x_{\epsilon}=1$ corresponds
to the scaling dimension of the energy operator, as in the $XX$ chain.

\section{Entanglement Entropy}

We are interested, now, in the finite-size corrections of the Rényi
entropy, in the critical regions I and II. The leading finite-size
corrections of Rényi entropy is related with the central charge, while
the sub-leading corrections are governed by the exponent $p_{\alpha}$.\citealp{entropyosc}
Below, we determine the central charge $c$ and the exponent $p_{\alpha}$
by using the universal behavior of these corrections.

In the case of quadratic Hamiltonians, as the one considered in this
work, the Rényi entropy can be evaluated through the correlation matrix
method\citep{0305-4470-36-14-101,PhysRevB.64.064412} (see also Refs.
\onlinecite{reviewriera} and \onlinecite{reviewpeschelBJP}). The
main idea of this method is that the eigenvalues $\{\nu_{j}\}$ of
the correlation matrix $C_{p,q}=<c_{p}^{\dagger}c_{q}>$, $p,q=1,...,\ell$;
are related with the eigenvalues of the reduced density matrix $\rho_{{\cal {A}}}$.\citep{0305-4470-36-14-101,PhysRevB.64.064412}
Due to this fact, the Rényi entropy can be expressed in terms of the
eigenvalues $\{\nu_{j}\}$ in the following form 
\begin{equation}
S_{\alpha}(L,\ell)=\frac{1}{1-\alpha}\sum_{j=1}^{\ell}\textrm{ln}\left[\nu_{j}^{\alpha}+\left(1-\nu_{j}\right)^{\alpha}\right]\textrm{.}\label{eq:renyieigen}
\end{equation}
We will present data only for $\ell$ even due to the geometry of
the two-leg ladder, although the odd sites also present the universal
behavior predicted by the CFT.

The elements of the correlation matrix $C_{p,q}$ associated with
the ground state wave function, can be easily evaluated, and are given
by\textcolor{red}{{} }
\[
\mathcal{C}_{(2p,2q)}=\frac{1}{2L}\frac{\sin[(p-q)2\pi\rho^{+}]+\sin[(p-q)2\pi\rho^{-}]}{\sin[\frac{(p-q)\pi}{L}]},
\]
\[
\mathcal{C}_{(2p-1,2q)}=-\frac{1}{2L}\sum_{k_{F}^{+}<|k|\leqslant k_{F}^{-}}\frac{J_{1}+J_{2}\exp(-ik)}{\sqrt{J_{1}^{2}+J_{2}^{2}+2J_{1}J_{2}\textrm{cos}k}}e^{-ik(p-q)},
\]
\begin{equation}
\mathcal{C}_{(2p-1,2q-1)}=\mathcal{C}_{(2p,2q)}\textrm{ and }\mathcal{C}_{(2p-1,2q)}=\mathcal{C}_{(2q,2p-1)}^{*}.\label{eq:correlmatrix}
\end{equation}
Similar expressions can also be obtained for the excited states.

In the next two subsections, we determine numerically the eigenvalues
of the correlation matrix $C_{p,q}$ in order to obtain the Rényi
entropy of the ground state as well as of the excited states.

\subsection{Ground State}

\begin{figure}
\begin{centering}
\psfrag{axis1}{\scalebox{1.5}{$S_1(L,\ell)$}}
\psfrag{axis2}{\scalebox{1.5}{$\ell$}}
\psfrag{axis3}{\scalebox{1.1}{$S_1(L,\ell)$}}
\psfrag{axis4}{\scalebox{1.2}{$\ell$}}
\includegraphics[scale=0.34]{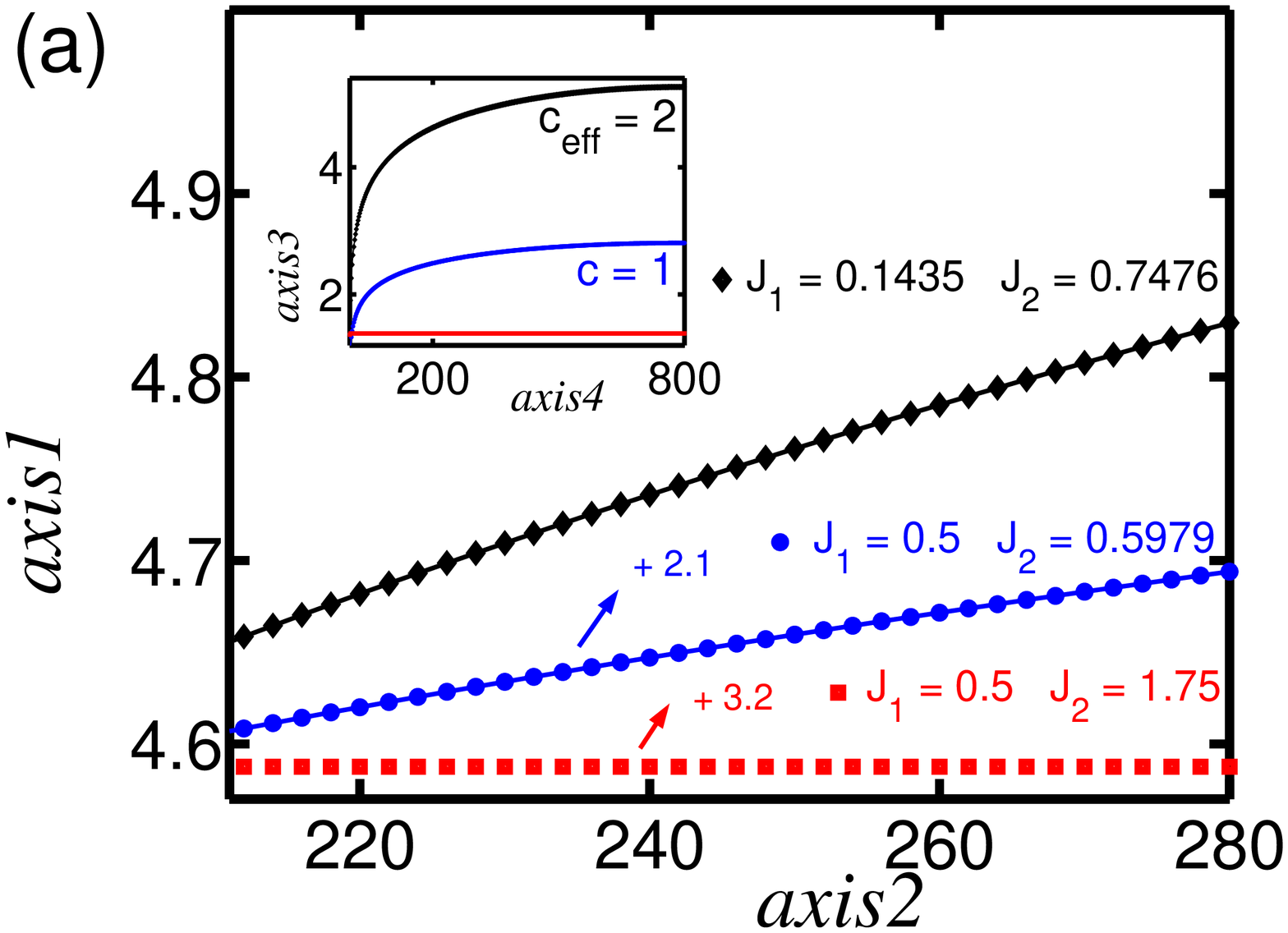}
\par\end{centering}

\begin{centering}
\psfrag{axis1}{\scalebox{1.5}{$D_1(L,\ell)$}}
\psfrag{axis2}{\scalebox{1.5}{$\ell$}}
\psfrag{axis3}{\scalebox{1.2}{$D_1(L,\ell)$}}
\psfrag{axis4}{\scalebox{1.2}{$\ell$}}
\includegraphics[scale=0.34]{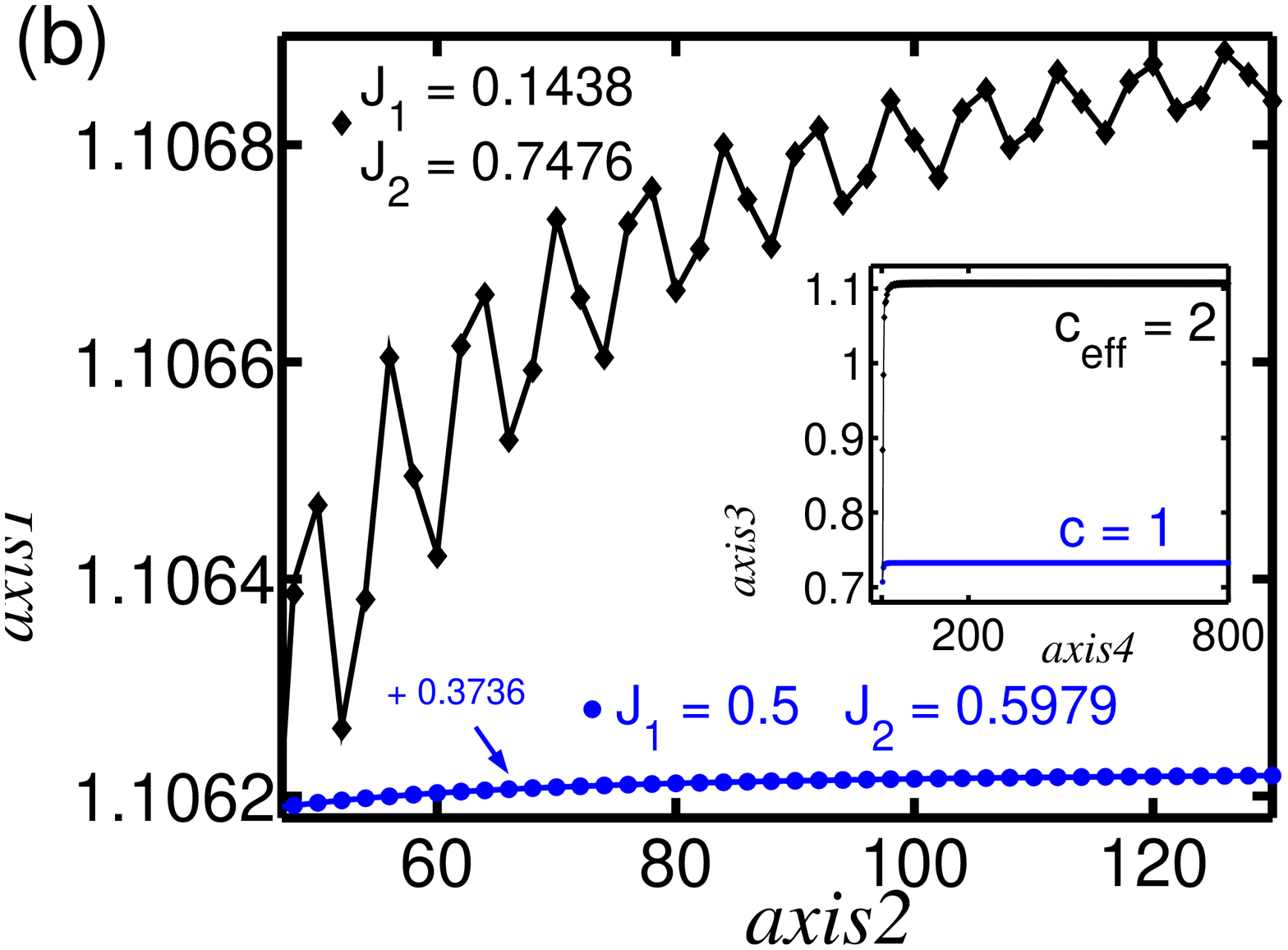}
\par\end{centering}

\caption{\label{fig4} (Color online). (a) The von Neumann entropy $S_{1}(L,\ell)$
of the ground state for systems of size $L=800$, $h=0$, $J_{3}=1$
and some values of $J_{1}$ and $J_{2}$ (see legend). The couplings
associate with the black, blue, and red data correspond to the couplings
belonging to the regions I, II and III, respectively {[}see Fig. (\ref{fig3}){]}.
(b) Results of the difference $D_{1}(L,\ell)$ for the same parameters
of figure (a). Only few sites are presented. The solid lines in these
figures connect the fitted points (see text). Insets: Show $S_{1}$
{[}figure (a){]} and $D_{1}$ {[}figure (b){]} for all sites. In order
to show all data in the figures, we added some constants in the values
of $S_{1}$ and $D_{1}$, as indicated by the arrows.}
\end{figure}

Let us consider first the Rényi entropy of the ground state. In Fig.
\ref{fig4}(a), we present the von Neumann entropy $S_{1},$ as function
of $\ell$ for systems of size $L=800$, $h=0$, and three sets of
couplings $(J_{1},J_{2},J_{3})$. These sets of couplings illustrate
the behavior of $S_{1}$ in the three distinct regions of the phase
diagram depicted in Fig. \ref{fig3}. As we observed in this figure,
in the gapped phase (region III) the von Neumann entropy tends to
a constant, as expected. On the other hand, for the critical regions
I and II, $S_{1}$ increases in agreement with the CFT prediction
{[}Eq. (\ref{eq:entropyCFT}){]}. The black/blue lines in Fig. \ref{fig4}(a)
are fits to our data using Eq. (\ref{eq:entropyCFT}). The central
charge $c$ obtained through this fit is $c=1.00001$ $(c_{eff}=2.00021)$
for the sets of couplings associated with the critical region II (I).
Similar results were acquired for several other sets of coupling for
these two distinct critical regions. These results show that for the
region II, the low-energy physics is describe by a CFT with central
charge $c=1$, as we have already predicted {[}see Eq. (\ref{eq:Egsfinite}){]}.
It is very interesting to note that in the region I, where we have two gapless
modes (instead of one as in the region II), we obtain an effective
central charge $c_{eff}=2$. This result resembles to the one found
for the finite-size corrections of the ground state energy {[}Eq.
(\ref{eq:Egsfinite}){]}. However, unlike the finite-size corrections
of the energies, the entropy does not depend on the sound velocities.
Then, it is expected that \emph{each gapless mode} contributes for
the finite-size corrections of the entropy as Eq. (\ref{eq:entropyCFT}).
For this reason, we get $c_{eff}=2c=2$. 

Another interesting feature of the von Neumann entropy in the region
I, is that it presents an unexpected oscillation.
In order to observe better these oscillations is convenient to define
the following difference

\[
D_{\alpha}(L,\ell)=S_{\alpha}(L,\ell)-\frac{c_{eff}}{6}\left(1+\frac{1}{\alpha}\right)
\ln\left[\frac{2L}{\pi}\sin\left(\frac{\pi\ell}{2L}\right)\right],
\]
where $c_{eff}=$1(2) in the region II(I).
From Eqs. (\ref{eq:renyientropy})-(\ref{eq:entropyUnusual}), it
is expected that the difference $D_{\alpha}$ behaves as

\[
D_{\alpha}(L,\ell)=c_{\alpha}+
\]

\begin{equation}
\frac{\left[a_{1}\delta_{1,\alpha}+g_{\alpha}(1-\delta_{1,\alpha})\cos(\kappa\ell+\phi)\right]}{L^{p_{\alpha}}}\left|\sin\left(\pi\frac{\ell}{2L}\right)\right|^{-p_{\alpha}}.\label{eq:Diffentropy}
\end{equation}
In Fig. \ref{fig4}(b), we show $D_{1}(L,\ell)$ for the two sets
of coupling parameters presented in Fig. \ref{fig4}(a).
If $J_1\ne J_2$, oscillations between the odd and
even sites appear naturally in the entropy. The oscillation
observed in Fig. \ref{fig4}(b), for the set of coupling belonging to the
Region I,  is not related with the "dimerization",
is present even for  $J_1=J_2$.
According to a previous conjecture of Xavier and
Alcaraz,\citep{XavierAlca2012} based in studies of several models
with just one gapless mode, the von Neumann entropy of single interval
should not present these oscillations. Our present results, show that
the same conjecture does not apply for models with more than one gapless
mode. These authors also conjectured that the sub-leading finite-size
correction of the von Neumann entropy decays with the exponent $p_{1}=\nu=2$.
Indeed, we also have observed this decay in the region II. The blue
solid line in Fig. \ref{fig4}(b) is the fit to our data assuming
that $D_{1}(L,\ell)$ behaves as Eq. (\ref{eq:Diffentropy}). The
exponent that we get by this fit is $p_{1}=2.0001$, in agreement
with the previous conjecture. 

In the region I, where we have two gapless mode, we fit our data assuming
that $D_{1}(L,\ell)$ has an oscillating term different of Eq. (\ref{eq:entropyUnusual}).
In this region, we had to replace $g_{1}(1-\delta_{1,\alpha})\cos\left(\kappa\ell+\phi\right)$
in Eq. (\ref{eq:entropyUnusual}) by the following term: $g_{1}\cos\left(\ell k_{F}^{-}\right)+g_{1}^{(2)}\cos\left(\ell k_{F}^{+}\right)+g_{1}^{(3)}\cos\left(\frac{k_{F}^{-}+k_{F}^{+}}{2}\ell\right)+g_{1}^{(4)}\cos\left(\frac{k_{F}^{-}-k_{F}^{+}}{2}\ell\right)$,
in order to fit our data. By using this oscillating term, we are able
to fit perfectly our data, as shown in Fig. \ref{fig4}(b). The exponent
obtained in this later case is $p_{1}=\nu=2.0007$. Similar results
were also observed for other couplings. The motivation for us consider
the above oscillating term is given below. These results indicate
that the sub-leading correction of the von Neumann entropy indeed
has the exponent $p_{1}=\nu=2$ even for models with more than one
gapless modes.

\begin{figure}
\begin{centering}
\psfrag{axis1}{\scalebox{1.5}{$S_3(L,\ell)$}}
\psfrag{axis2}{\scalebox{1.5}{$\ell$}}
\psfrag{axis3}{\scalebox{1.2}{$S_{10}(L,\ell)$}}
\psfrag{axis4}{\scalebox{1.2}{$\ell$}}
\includegraphics[scale=0.34]{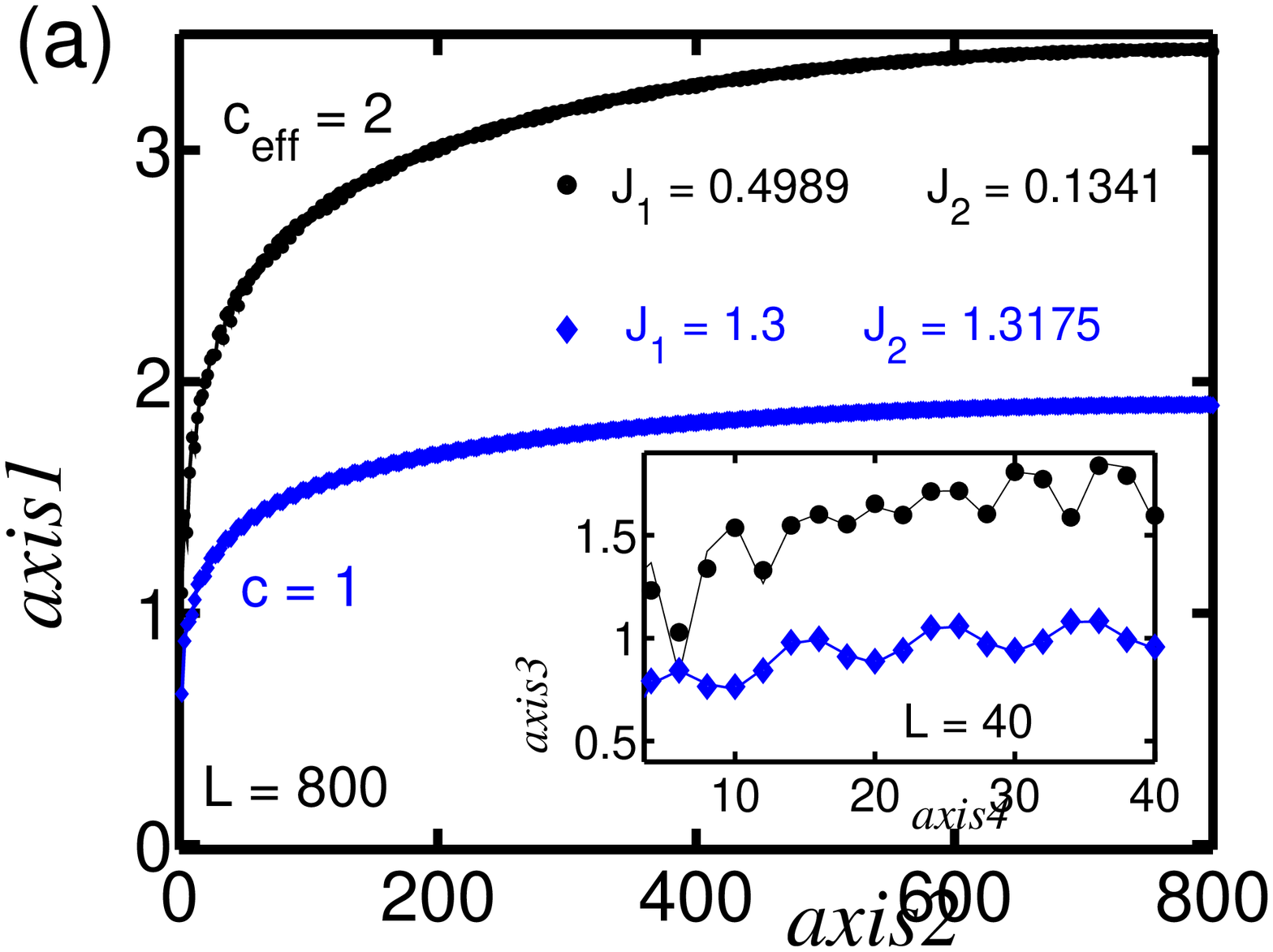}
\par\end{centering}

\begin{centering}
\psfrag{axis1}{\scalebox{1.5}{$D_3(L,\ell)$}}
\psfrag{axis2}{\scalebox{1.5}{$\ell$}}
\includegraphics[scale=0.34]{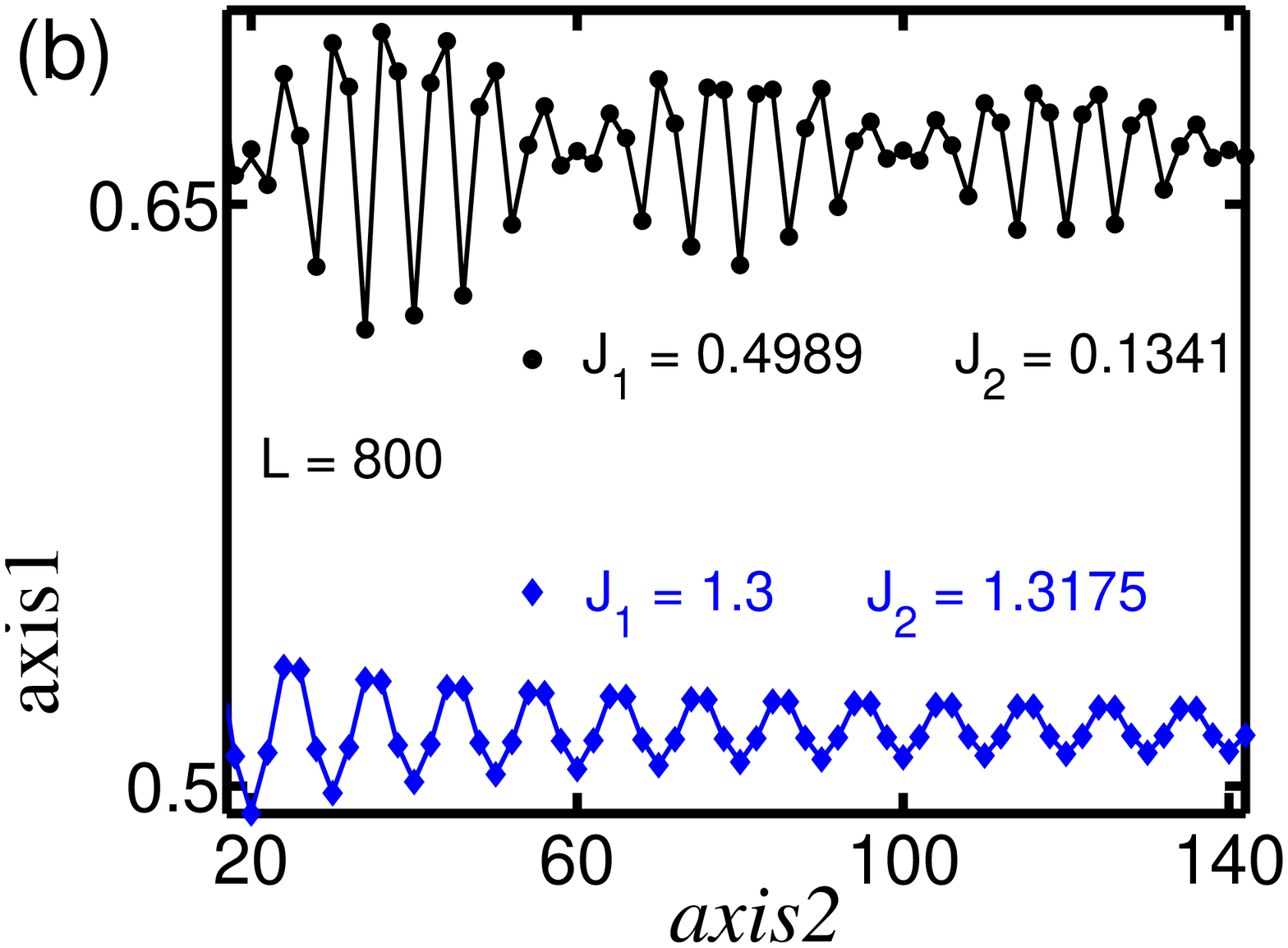}
\par\end{centering}

\caption{\label{fig5} (Color online). (a) The Rényi entropy of the ground
state for systems of size $L=800$, $h=0$, $J_{3}=1$, $\alpha=3$,
and two values of $J_{1}$ and $J_{2}$ (see legend). The couplings
associate with the black and blue lines, correspond to couplings belonging
to the regions I, and II, respectively {[}see Fig. (\ref{fig3}){]}.
Inset: Results for $L=40$. (b) The difference
$D_{\alpha}(L,\ell)$ for the same parameters of figure (a). Only
few sites are presented. The symbols in the figures (a) and (b) are the
numerical data and the solid lines connect the fitted points (see
text).}
\end{figure}

Let us now consider the $\alpha-$Rényi entropies of the ground state
with $\alpha>1$. As illustration, we present in Fig. \ref{fig5}(a),
$S_{\alpha}(L,\ell)$ as function of $\ell$ for systems of size $L=800$
and $L=40$, two values of $\alpha$, and some couplings. In the case
that $\alpha>1$, it is expected oscillations in the entropy even
for systems with PBC,\citep{entropyosc} as mentioned in the introduction.
Moreover, it is expected that the amplitude of the oscillations decrease,
as the system size increases {[}see Eq. (\ref{eq:entropyUnusual}){]}.
Indeed, this feature is clearly noted in a system with smaller size,
as depicted in the inset of Fig. \ref{fig5}(a). 

Once again, in order to observe better these oscillations we investigate
the difference $D_{\alpha}(L,\ell)$. In Fig. \ref{fig5}(b), we show
$D_{\alpha}(L,\ell)$ vs $\ell$ for the same data presented in Fig.
\ref{fig5}(a) with $\alpha=3$. Let us first discuss the data of
the Rényi entropy in the region II. In this region, we are able to
fit our data by using Eq. (\ref{eq:entropyUnusual}) with $\kappa=k_{F}^{-}$
and $\phi=0$. The exponent we get from this fit is $p_{3}=0.666$.
Similar fits of $S_{\alpha}$ (not shown) for $\alpha=2$ and $\alpha=4$
give $p_{2}=0.999$ and $p_{4}=0.497$, respectively. These results
indicate that $p_{\alpha}=2/\alpha$, as in the $XX$ chain. 

On the other hand, it is not possible to fit the data of the Rényi
entropy, in the region I, if we consider that the oscillating term
behaves just as $\cos(\kappa\ell+\phi)$, as it was done in the region
II. In order to get some insight about these oscillations in the region
I, we calculate analytically the spin-spin correlation function $\left\langle s_{2n}^{z}s_{2n+\mathbb{\mathtt{2\ell}}}^{z}\right\rangle $,
which is given by 
\[
\left\langle s_{2n}^{z}s_{2(n+\mathbb{\mathtt{\ell)}}}^{z}\right\rangle =m^{2}-\frac{1}{4L^{2}\sin^{2}(\pi\ell/2L)}\Bigg[1-\frac{1}{2}\cos\left(\ell k_{F}^{-}\right)-
\]
\[
\frac{1}{2}\cos\left(\ell k_{F}^{+}\right)-\cos\left(\frac{k_{F}^{-}+k_{F}^{+}}{2}\ell\right)+\cos\left(\frac{k_{F}^{-}-k_{F}^{+}}{2}\ell\right)\Bigg].
\]
Motivated by the fact  that it is expected that the oscillations of the entropy are associated
with the anti-ferromagnetic nature of the Hamiltonian,\citealp{entropyaffleckosc}
we assume that the difference $D_{\alpha}(L,\ell)$, in the region
I, behaves as
\[
D_{\alpha}(L,\ell)=c_{\alpha}+\Bigg[a_{1}\delta_{1,\alpha}+g_{\alpha}\cos\left(\ell k_{F}^{-}\right)+g_{\alpha}^{(2)}\cos\left(\ell k_{F}^{+}\right)+
\]

\begin{equation}
g_{\alpha}^{(3)}\cos\left(\frac{k_{F}^{-}+k_{F}^{+}}{2}\ell\right)+g_{\alpha}^{(4)}\cos\left(\frac{k_{F}^{-}-k_{F}^{+}}{2}\ell\right)\Bigg]\left|L\sin\left(\pi\frac{\ell}{2L}\right)\right|^{-p_{\alpha}}.\label{eq:Diffentropy-b}
\end{equation}
Indeed, as illustrated in Fig. \ref{fig5}(b), we obtain a very nice
fit of our data if we assume that $D_{\alpha}$ behaves as the above
equation. The exponent we get from the fit is $p_{3}=0.667$. 

\begin{table}
\begin{center}
\begin{tabular}{lrr}
{\scriptsize $\left\{ \hspace{-2mm}\begin{array}{l}
\circ\circ|\bullet\bullet\cdots\bullet\bullet|\circ\circ\\
\circ\circ|\bullet\bullet\cdots\bullet\bullet|\circ\circ
\end{array}\hspace{-2mm}\right\} $\vspace{1mm}} & \multicolumn{2}{l}{gs - Region (I)}\tabularnewline
{\scriptsize $\left\{ \hspace{-2mm}\begin{array}{l}
\circ\circ{\color{white}|}\circ\circ\cdots\circ\circ{\color{white}|}\circ\circ\\
\circ\circ|\bullet\bullet\cdots\bullet\bullet|\circ\circ
\end{array}\hspace{-2mm}\right\} $\vspace{2mm}} & \multicolumn{2}{l}{gs - Region (II)}\tabularnewline
\multicolumn{3}{l}{Compact Excitations{\scriptsize \vspace{1.5mm}}}\tabularnewline
{\scriptsize $\left\{ \hspace{-2mm}\begin{array}{l}
\circ\circ{\color{white}|}\circ\circ\cdots\circ\circ{\color{white}|}\circ\circ\\
\circ\bullet|\bullet\bullet\cdots\bullet\bullet|\bullet\circ
\end{array}\hspace{-2mm}\right\} $\vspace{1mm}} & \textcolor{black}{\scriptsize $(:)_{+}(:\pm1)_{-}$} & \textcolor{black}{\scriptsize $I_{-}=2$}\tabularnewline
{\scriptsize $\left\{ \hspace{-2mm}\begin{array}{l}
\circ\circ{\color{white}|}\circ\circ\cdots\circ\circ{\color{white}|}\circ\circ\\
\circ\circ|\bullet\bullet\cdots\bullet\bullet|\bullet\circ
\end{array}\hspace{-2mm}\right\} $\vspace{1mm}} & \textcolor{black}{\scriptsize $(:)_{+}(:1)_{-}$} & \textcolor{black}{\scriptsize $I_{-}=1$}\tabularnewline
{\scriptsize $\left\{ \hspace{-2mm}\begin{array}{l}
\circ\circ|\bullet\bullet\cdots\bullet\bullet|\bullet\circ\\
\circ\circ|\circ\bullet\cdots\bullet\bullet|\circ\circ
\end{array}\hspace{-2mm}\right\} $\vspace{1mm}} & \textcolor{black}{\scriptsize $(:1)_{+}(-1:)_{-}$} & \textcolor{black}{\scriptsize $I_{\pm}=\pm1$}\tabularnewline
{\scriptsize $\left\{ \hspace{-2mm}\begin{array}{l}
\circ|\bullet\cdots\bullet\circ\circ|\circ\circ\circ\\
\circ|\bullet\cdots\bullet\bullet\bullet|\bullet\bullet\circ
\end{array}\hspace{-2mm}\right\} $\vspace{2mm}} & \textcolor{black}{\scriptsize $(1,2:)_{+}(:1,2)_{-}$} & \textcolor{black}{\scriptsize $I_{\pm}=\mp2$ $Q_{\pm}=\mp1$}\tabularnewline
\multicolumn{3}{l}{Non-Compact Excitations{\scriptsize \vspace{1.5mm}}}\tabularnewline
{\scriptsize $\left\{ \hspace{-2mm}\begin{array}{l}
\circ\circ|\bullet\bullet\cdots\bullet\bullet|\circ\circ\\
\circ\circ|\bullet\bullet\cdots\bullet\circ|\bullet\circ
\end{array}\hspace{-2mm}\right\} $\vspace{1mm}} & \textcolor{black}{\scriptsize $(:)_{+}(1:1)_{-}$} & \textcolor{black}{\scriptsize $j_{-}=1$}\tabularnewline
{\scriptsize $\left\{ \hspace{-2mm}\begin{array}{l}
\circ\bullet|\circ\bullet\cdots\bullet\bullet|\circ\circ\\
\circ\bullet|\circ\bullet\cdots\bullet\bullet|\circ\circ
\end{array}\hspace{-2mm}\right\} $\vspace{1mm}} & \textcolor{black}{\scriptsize $(-1:-1)_{\pm}$} & \textcolor{black}{\scriptsize $j\prime_{\pm}=1$}\tabularnewline
{\scriptsize $\left\{ \hspace{-2mm}\begin{array}{l}
\circ{\color{white}|}\circ\cdots\circ\circ\circ{\color{white}|}\circ\circ\circ\\
\circ|\bullet\cdots\bullet\circ\circ|\bullet\bullet\circ
\end{array}\hspace{-2mm}\right\} $\vspace{1mm}} & \textcolor{black}{\scriptsize $(:)_{+}(1,2:1,2)_{-}$} & \textcolor{black}{\scriptsize $j_{-}=4$}\tabularnewline
{\scriptsize $\left\{ \hspace{-2mm}\begin{array}{l}
\circ{\color{white}|}\circ\circ\cdots\circ\circ\circ{\color{white}|}\circ\circ\\
\circ|\circ\bullet\cdots\bullet\bullet\circ|\bullet\circ
\end{array}\hspace{-2mm}\right\} $} & \textcolor{black}{\scriptsize $(:)_+(\pm1:1)_{-}$} & \textcolor{black}{\scriptsize $I_{-}=j_{-}=1$}\tabularnewline
\end{tabular}
\par\end{center}

\caption{Schematic representations of some states. The open (closed) cycles
are the empty (occupied) levels. The two rows of cycles inside the
symbol $\left\{ \right\} $ denotes the two branches, while the vertical
lines show the positions of the Fermi momenta. We also present the
excitations in terms of $I_{\sigma}$ and $Q_{\sigma}$ [see
Eq. (\ref{eq:exctfinite})],
and the notation used by the authors of
Ref. \onlinecite{chicoentropyext2}. In the later case
$(h_1h_2...:p_1p_2...)_\sigma$ represents an excited state with holes
(particles) in the $h_i$s ($p_i$s) allowed momentum values  below (above) the
Fermi point of the branch $\sigma$.}
\end{table}

Similar results, as the ones presented above, were obtained for several
other values of $m$ and coupling parameters. In addition to that,
the results for some values of $\alpha$ strongly support that the exponent
$p_{\alpha}=2/\alpha$, as happens in the $XX$ chain.\citealp{entropyosc}
These results are in agreement with the conjecture\citealp{XavierAlca2012}
that the exponent $p_{\alpha}$ is related with the dimension of the
energy operator $x_{\epsilon}$ by $p_{\alpha}=2x_{\epsilon}/\alpha$,
which is $x_{\epsilon}=1$ in the present model. 

\begin{figure}
\begin{centering}
\psfrag{axis1}{\scalebox{1.5}{$\Delta S_{2}(L,\ell)$}}
\psfrag{axis2}{\scalebox{1.5}{$\ell$}}
\psfrag{axis3}{\scalebox{1.1}{$\Delta S_{2}(L,\ell)$}}
\psfrag{axis4}{\scalebox{1.2}{$\ell$}}
\psfrag{excA}{\scalebox{0.7}{$\left\{
\begin{array}{c}\left(\pm1:\pm1\right)_{\pm}\\n=4\end{array}\right.$}}
\psfrag{excB}{\scalebox{0.7}{$\left\{
\begin{array}{c}(-1:-1)_{+}(\pm1:\pm1)_{-}\\n=3\end{array}\right.$}}
%
\psfrag{excC}{\scalebox{0.7}{$\left\{ \begin{array}{c}(1:1)_{\pm}\\n=2\end{array} \right.$}} 
\psfrag{excD}{\scalebox{0.7}{$\left\{
\begin{array}{c}\left(:\right)_{+}\left(1:1\right)_{-}\\n=1\end{array}\right.$}}
\includegraphics[scale=0.34]{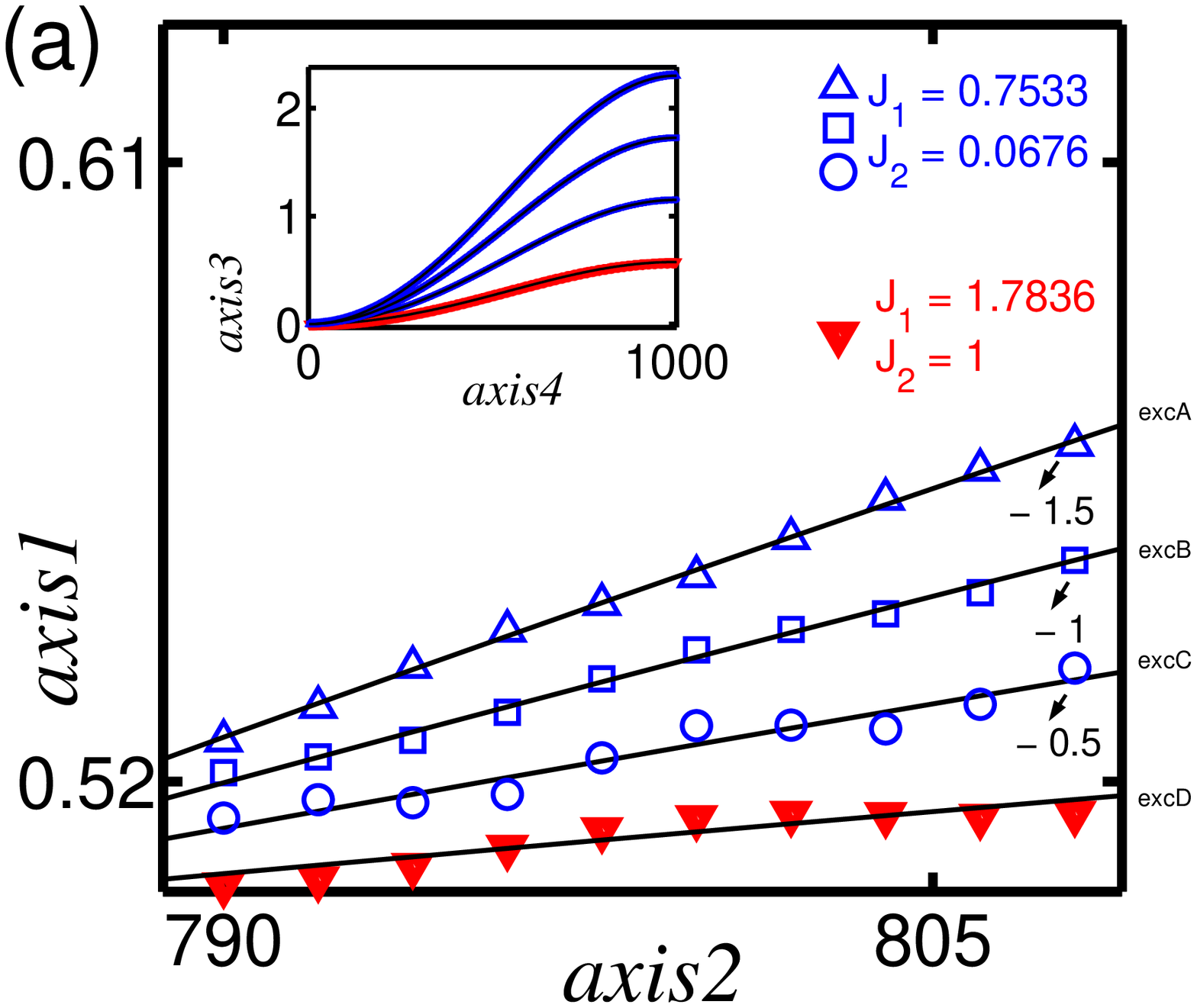}
\par\end{centering}

\begin{centering}
\psfrag{axis1}{\scalebox{1.5}{$\Delta S_{1}(L,\ell)$}}
\psfrag{axis2}{\scalebox{1.5}{$\frac{2\pi^2}{3}(\frac{\ell}{2L})^2$}}
\includegraphics[scale=0.34]{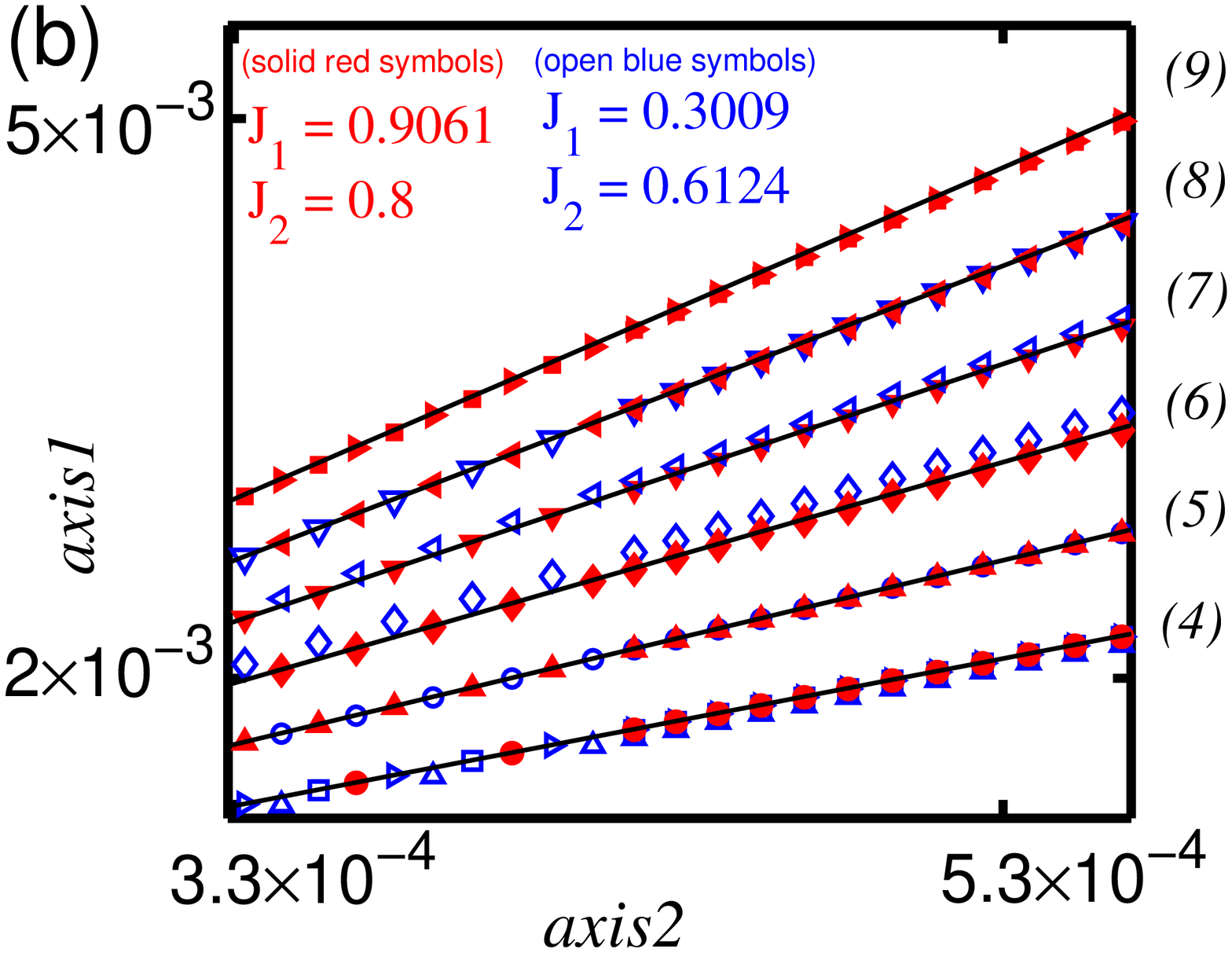}
\par\end{centering}

\caption{\label{fig6} (Color online). (a) \textbf{$\Delta S_{2}$} vs $\ell$
for some particle-hole excitations for systems of size $L=1000$,
$h=0$, and some couplings (see legend). The symbols are the numerical
data and the solid lines connect the fitted data (see text). The
solid curves from the top to down have values of $n$: 4,3,2 and 1
{[}see  Eq. \ref{eq:fitF2}{]}. The blue(red) symbols are data from
the region I(II). Only few sites are presented. In order
to show all data in the figure, we added some constants in the values
of $\Delta S_2$, as indicated by the arrows. We also present the
excitations in terms of the notation of Ref. \onlinecite{chicoentropyext2}.
Inset: $\Delta S_{2}$ for all sites. (b) \textbf{$\Delta S_{1}$} vs
$\frac{2\pi^2}{3}(\ell/2L)^2$ for several excitations for systems of size
$L=10000$ (see Table II). }
 
\end{figure}

\subsection{Excited states}

Finally, we consider the Rényi entropy of the excited states. Differently
of the Rényi entropy of the ground state, where several studies indeed
confirmed the universal behavior predicted by the CFT, few works considered
the excited states.\citealp{chicoentropy,chicoentropyext2,CFThomog,calabresecont,CFTexcPRA,chicosarandy,exctdalmonte,unsu2012}
Certainly, more studies are highly desired in this front and in this
subsection we intent to provided a study of the Rényi entropy of several
excited states of the Hamiltonian (\ref{Eq:Hamiltonian}).

First, we consider the Rényi entropy associated with compact excitations.
The compact excitation are those that do not present holes in the momentum
space. In Table I, we depict few examples of such excitations. We
found that $\Delta S_{\alpha}(L,\ell)=S_{\alpha}^{exct}-S_{\alpha}^{gs}\lesssim10^{-3}$
for these compact excitations, as well as many others not presented
in this table. These results are in agreement with the prediction, done
by the authors of the Refs. \onlinecite{chicoentropy} and \onlinecite{chicoentropyext2},
that $S_{\alpha}^{exc}=S_{\alpha}^{gs}$ for the compact excitations. 

The Rényi entropies associated with the non-compact excitations are,
in general, different from the ground state. Let us consider  particle-hole
excitations in the right/left Fermi points of the branch $\sigma$.
In Fig. \ref{fig6}(a), we present $\Delta S_{2}$ as function of $\ell$
for some these excitations for systems with $2L$ sites. 
As observed in this figure, the Rényi entropies associated with these
non-compact excitations increase with $\ell$. Moreover our results,
show that the  $\Delta S_{2}$ data can be perfectly fitted (except by
the unusual oscillations) if we
assume that {[}see Eq.(\ref{eq:F2}){]}

\begin{equation}
\Delta S_{2}=-nF_{i\partial\phi}^{(2)}=-n\ln\left(1-2s^{2}+3s^{4}-2s^{6}+8s^{8}\right),\label{eq:fitF2}
\end{equation}
where $s=\sin(\pi\ell/4L)$ and $n$ is the number of particle-hole excitations associated with
the excited state.

In the case of just \emph{one gapless} mode, the Eq. (\ref{eq:fitF2})
was determined analytically for the free boson theory.\citep{chicoentropyext2}
Note that for a free boson theory with \emph{two decoupled field $\phi^{\sigma}$
($\sigma=\pm)$} the Eq. (\ref{eq:fitF2}) is  valid.  

Finally, we consider $\Delta S_{1}(L,\ell)=S_{1}^{exc}-S_{1}^{gs}$
for several excitations. In particular we focus on the behavior of
$\Delta S_{1}$ in the regime that $(\ell/L)\ll1$. In this regime
it is expected that $\Delta S_{1}$, for a system with $2L$ sites, behaves
as\citealp{chicoentropy,chicoentropyext2}  
\begin{equation}
\Delta S_{1}=\frac{2\pi^{2}}{3}d^{\beta}\left(\frac{\ell}{2L}\right)^{2}+o[(\ell/L)^{2}],\label{eq:detas1}
\end{equation}
where $d^{\beta}$ is the scaling dimension of the operator assocociate
with the excitation. This behavior of $\Delta S_{1}$ above was proofed\citealp{chicoentropy}
in the case of the primay operator $i\partial\phi$
 and verified numerically
for the descendant operators.\citealp{chicoentropyext2}

The scaling dimensions of
the $XX$ chain have the same structure of the Gaussian model\citealp{critical3} (with
compactification radius $R=2)$ and are given by $d_{I,Q,j,j'}=x_{I,Q}+j+j'$,
where $x_{I,Q}=(\frac{I^{2}}{4}+Q^{2})$ are the scaling dimensions of the
primary operators and $j,j'=0,1,2...$ {[}see Eq. (\ref{eq:2}){]}. 
Note that
the energy associated with an operator with dimension $d_{I,Q,j,j'}$ (in the
Gaussian model)
has momentum $\Delta P_{I,Q}^{j,j'}=\frac{2\pi}{L}(IQ+j-j')$.\citealp{critical3}
We also found a similar result. The momentum of a state associated with the
scaling dimension $\sum_{\sigma=\pm}(d^\sigma_{I_\sigma,Q_\sigma}+j_\sigma+j_\sigma')$  is given by
$\Delta
P=\sum_{\sigma\pm}\Delta P_{I_\sigma,Q_\sigma,j_\sigma,j_\sigma'}^\sigma$, 
where
$\Delta
P_{I_\sigma,Q_\sigma,j_\sigma,j_\sigma'}^\sigma=\frac{2\pi}{L}\left[I_\sigma
  (Q_\sigma+\delta_\alpha/2)+j_\sigma-j_\sigma'\right]$. The  momenta of 
the configurations presented in Table II are in perfect agreement with
the values of $\Delta P$.


\begin{table}
\noindent \begin{center}
REGION II
\begin{tabular}{rcccc}
 & {\footnotesize $I_{-}$} & {\footnotesize $Q_{-}$} & {\footnotesize $j_{-}$} & \tabularnewline
\textcolor{red}{\Large $\bullet$}{\scriptsize $\left\{ \hspace{-2mm}\begin{array}{l}
\circ{\color{white}|}\hspace{-0.5mm}\circ\hspace{-0.5mm}\cdots\hspace{-0.5mm}\circ\hspace{-0.5mm}\circ{\color{white}|}\hspace{-0.5mm}\circ\hspace{-0.5mm}\circ\hspace{-0.5mm}\circ\hspace{-0.5mm}\circ\hspace{-0.5mm}\circ\hspace{-0.5mm}\circ\circ\\
\circ|\hspace{-0.5mm}\bullet\hspace{-0.5mm}\cdots\hspace{-0.5mm}\bullet\hspace{-0.5mm}\circ|\hspace{-0.5mm}\circ\hspace{-0.5mm}\circ\hspace{-0.5mm}\circ\hspace{-0.5mm}\bullet\hspace{-0.5mm}\circ\hspace{-0.5mm}\circ\circ
\end{array}\hspace{-2mm}\right\} $\vspace{1mm}} & {\footnotesize $0$} & {\footnotesize $0$} & {\footnotesize $4$} & \textcolor{black}{\scriptsize $(:)_{+}(1:4)_{-}$}\tabularnewline
\textcolor{red}{\large ${\color{red}\blacktriangle}$}{\scriptsize $\left\{ \hspace{-2mm}\begin{array}{l}
\circ{\color{white}|}\hspace{-0.5mm}\circ\hspace{-0.5mm}\cdots\hspace{-0.5mm}\circ\hspace{-0.5mm}\circ{\color{white}|}\hspace{-0.5mm}\circ\hspace{-0.5mm}\circ\hspace{-0.5mm}\circ\hspace{-0.5mm}\circ\hspace{-0.5mm}\circ\hspace{-0.5mm}\circ\circ\\
\circ|\hspace{-0.5mm}\bullet\hspace{-0.5mm}\cdots\hspace{-0.5mm}\bullet\hspace{-0.5mm}\circ|\hspace{-0.5mm}\circ\hspace{-0.5mm}\circ\hspace{-0.5mm}\circ\hspace{-0.5mm}\circ\hspace{-0.5mm}\bullet\hspace{-0.5mm}\circ\circ
\end{array}\hspace{-2mm}\right\} $\vspace{1mm}} & {\footnotesize $0$} & {\footnotesize $0$} & {\footnotesize $5$} & \textcolor{black}{\scriptsize $(:)_{+}(1:5)_{-}$}\tabularnewline
\textcolor{red}{${\color{red}\blacklozenge}$}{\scriptsize $\left\{ \hspace{-2mm}\begin{array}{l}
\circ{\color{white}|}\hspace{-0.5mm}\circ\hspace{-0.5mm}\cdots\hspace{-0.5mm}\circ\hspace{-0.5mm}\circ{\color{white}|}\hspace{-0.5mm}\circ\hspace{-0.5mm}\circ\hspace{-0.5mm}\circ\hspace{-0.5mm}\circ\hspace{-0.5mm}\circ\hspace{-0.5mm}\circ\circ\\
\circ|\hspace{-0.5mm}\bullet\hspace{-0.5mm}\cdots\hspace{-0.5mm}\bullet\hspace{-0.5mm}\circ|\hspace{-0.5mm}\circ\hspace{-0.5mm}\circ\hspace{-0.5mm}\circ\hspace{-0.5mm}\circ\hspace{-0.5mm}\circ\hspace{-0.5mm}\bullet\circ
\end{array}\hspace{-2mm}\right\} $\vspace{1mm}} & {\footnotesize $0$} & {\footnotesize $0$} & {\footnotesize $6$} & \textcolor{black}{\scriptsize $(:)_{+}(1:6)_{-}$}\tabularnewline
{\scriptsize ${\color{red}\blacksquare}$$\left\{ \hspace{-2mm}\begin{array}{l}
\circ{\color{white}|}\hspace{-0.5mm}\circ\hspace{-0.5mm}\cdots\hspace{-0.5mm}\circ\hspace{-0.5mm}\circ\hspace{-0.5mm}\circ\hspace{-0.5mm}\circ{\color{white}|}\hspace{-0.5mm}\circ\hspace{-0.5mm}\circ\hspace{-0.5mm}\circ\hspace{-0.5mm}\circ\circ\\
\circ|\hspace{-0.5mm}\bullet\hspace{-0.5mm}\cdots\hspace{-0.5mm}\bullet\hspace{-0.5mm}\circ\hspace{-0.5mm}\circ\hspace{-0.5mm}\circ|\hspace{-0.5mm}\bullet\hspace{-0.5mm}\bullet\hspace{-0.5mm}\bullet\hspace{-0.5mm}\circ\circ
\end{array}\hspace{-2mm}\right\} $\vspace{1mm}} & {\footnotesize $0$} & {\footnotesize $0$} & {\footnotesize $9$} & \textcolor{black}{\scriptsize $(:)_{+}(1,2,3:1,2,3)_{-}$}\tabularnewline
\textcolor{red}{\large ${\color{red}\blacktriangleleft}$}{\scriptsize $\left\{ \hspace{-2mm}\begin{array}{l}
\circ{\color{white}|}\hspace{-0.5mm}\circ\hspace{-0.5mm}\cdots\hspace{-0.5mm}\circ\hspace{-0.5mm}\circ\hspace{-0.5mm}\circ\hspace{-0.5mm}\circ{\color{white}|}\hspace{-0.5mm}\circ\hspace{-0.5mm}\circ\hspace{-0.5mm}\circ\hspace{-0.5mm}\circ\circ\\
\circ|\hspace{-0.5mm}\bullet\hspace{-0.5mm}\cdots\hspace{-0.5mm}\bullet\hspace{-0.5mm}\bullet\hspace{-0.5mm}\circ\hspace{-0.5mm}\circ|\hspace{-0.5mm}\bullet\hspace{-0.5mm}\bullet\hspace{-0.5mm}\bullet\hspace{-0.5mm}\bullet\circ
\end{array}\hspace{-2mm}\right\} $\vspace{1mm}} & {\footnotesize $2$} & {\footnotesize $1$} & {\footnotesize $8$} & \textcolor{black}{\scriptsize $(:)_{+}(1,2:1,2,3,4)_{-}$}\tabularnewline
\textcolor{red}{\large ${\color{red}\blacktriangleright}$}{\scriptsize $\left\{ \hspace{-2mm}\begin{array}{l}
\circ\hspace{-0.5mm}\left.\,\circ\hspace{-0.5mm}\cdots\hspace{-0.5mm}\circ\hspace{-0.5mm}\circ\circ\right.\circ\hspace{-0.5mm}\circ\hspace{-0.5mm}\circ\hspace{-0.5mm}\circ\hspace{-0.5mm}\circ\hspace{-0.5mm}\circ\\
\circ\hspace{-0.5mm}\left|\bullet\hspace{-0.5mm}\cdots\hspace{-0.5mm}\bullet\hspace{-0.5mm}\circ\bullet\right|\hspace{-0.5mm}\circ\hspace{-0.5mm}\circ\hspace{-0.5mm}\circ\hspace{-0.5mm}\bullet\hspace{-0.5mm}\bullet\hspace{-0.5mm}\circ
\end{array}\hspace{-2mm}\right\} $\vspace{1mm}} & {\footnotesize $1$} & {\footnotesize $0$} & {\footnotesize $9$} & \textcolor{black}{\scriptsize $(:)_{+}(2:4,5)_{-}$}\tabularnewline
\textcolor{red}{\large ${\color{red}\blacktriangledown}$}{\scriptsize $\left\{ \hspace{-2mm}\begin{array}{l}
\circ{\color{white}|}\hspace{-0.5mm}\circ\hspace{-0.5mm}\cdots\hspace{-0.5mm}\circ\hspace{-0.5mm}\circ\hspace{-0.5mm}\circ\hspace{-0.5mm}\circ{\color{white}|}\hspace{-0.5mm}\circ\hspace{-0.5mm}\circ\hspace{-0.5mm}\circ\hspace{-0.5mm}\circ\circ\\
\circ|\hspace{-0.5mm}\bullet\hspace{-0.5mm}\cdots\hspace{-0.5mm}\bullet\hspace{-0.5mm}\circ\hspace{-0.5mm}\bullet\hspace{-0.5mm}\circ|\hspace{-0.5mm}\bullet\hspace{-0.5mm}\bullet\hspace{-0.5mm}\bullet\hspace{-0.5mm}\circ\circ
\end{array}\hspace{-2mm}\right\} $\vspace{1mm}} & {\footnotesize $1$} & {\footnotesize $0$} & {\footnotesize $7$} & \textcolor{black}{\scriptsize $(:)_{+}(1,3:1,2,3)_{-}$}\tabularnewline
\end{tabular}

REGION I

\begin{tabular}{rccccccc}
 & {\footnotesize $I_{+}$} & {\footnotesize $Q_{+}$} & {\footnotesize $j_{+}$} & {\footnotesize $I_{-}$} & {\footnotesize $Q_{-}$} & {\footnotesize $j_{-}$} & \tabularnewline
{\scriptsize ${\color{blue}{\color{blue}\square}}$$\left\{ \hspace{-2mm}\begin{array}{l}
\circ|\bullet\cdots\bullet\circ\circ|\bullet\bullet\circ\\
\circ|\bullet\cdots\bullet\bullet\bullet|\circ\circ\circ
\end{array}\hspace{-2mm}\right\} $\vspace{1mm}} & {\footnotesize $0$} & {\footnotesize $0$} & {\footnotesize $4$} & {\footnotesize $0$} & {\footnotesize $0$} & {\footnotesize $0$} & \textcolor{black}{\scriptsize $(1,2:1,2)_{+}(:)_{-}$}\tabularnewline
\textcolor{blue}{\large ${\color{blue}\triangleright}$}{\scriptsize $\left\{ \hspace{-2mm}\begin{array}{l}
\circ|\bullet\cdots\bullet\bullet\bullet|\circ\circ\circ\\
\circ|\bullet\cdots\bullet\circ\circ|\bullet\bullet\circ
\end{array}\hspace{-2mm}\right\} $\vspace{1mm}} & {\footnotesize $0$} & {\footnotesize $0$} & {\footnotesize $0$} & {\footnotesize $0$} & {\footnotesize $0$} & {\footnotesize $4$} & \textcolor{black}{\scriptsize $(:)_{+}(1,2:1,2)_{-}$}\tabularnewline
\textcolor{blue}{\large ${\color{blue}\triangledown}$}{\scriptsize $\left\{ \hspace{-2mm}\begin{array}{l}
\circ|\bullet\cdots\bullet\circ\circ|\bullet\bullet\circ\\
\circ|\bullet\cdots\bullet\circ\circ|\bullet\bullet\circ
\end{array}\hspace{-2mm}\right\} $\vspace{1mm}} & {\footnotesize $0$} & {\footnotesize $0$} & {\footnotesize $4$} & {\footnotesize $0$} & {\footnotesize $0$} & {\footnotesize $4$} & \textcolor{black}{\scriptsize $(1,2:1,2)_{\pm}$}\tabularnewline
\textcolor{blue}{\Large ${\color{blue}\circ}$}{\scriptsize $\left\{ \hspace{-2mm}\begin{array}{c}
\circ|\bullet\cdots\bullet\bullet\circ|\bullet\circ\circ\\
\circ|\bullet\cdots\bullet\circ\circ|\bullet\bullet\circ
\end{array}\hspace{-2mm}\right\} $\vspace{1mm}} & {\footnotesize $0$} & {\footnotesize $0$} & {\footnotesize $1$} & {\footnotesize $0$} & {\footnotesize $0$} & {\footnotesize $4$} & \textcolor{black}{\scriptsize $(1:1)_{+}(1,2:1,2)_{-}$}\tabularnewline
\textcolor{blue}{\large ${\color{blue}{\color{blue}\vartriangle}}$}{\scriptsize $\left\{ \hspace{-2mm}\begin{array}{c}
\circ|\bullet\cdots\bullet\circ\circ|\bullet\circ\circ\\
\circ|\bullet\cdots\bullet\bullet\circ|\bullet\bullet\circ
\end{array}\hspace{-2mm}\right\} $\vspace{1mm}} & {\footnotesize $1$} & {\footnotesize $0$} & {\footnotesize $2$} & {\footnotesize $-1$} & {\footnotesize $0$} & {\footnotesize $2$} & \textcolor{black}{\scriptsize $(1,2:1)_{+}(1:1,2)_{-}$}\tabularnewline
\textcolor{blue}{${\color{blue}\Diamond}$}{\scriptsize $\left\{ \hspace{-2mm}\begin{array}{c}
\circ\hspace{-0.5mm}\left|\bullet\cdots\bullet\hspace{-0.5mm}\bullet\hspace{-0.5mm}\bullet\hspace{-0.5mm}\circ\right|\hspace{-0.5mm}\bullet\hspace{-0.5mm}\bullet\hspace{-0.5mm}\bullet\hspace{-0.5mm}\circ\\
\circ\hspace{-0.5mm}\left|\bullet\cdots\bullet\hspace{-0.5mm}\circ\hspace{-0.5mm}\circ\hspace{-0.5mm}\circ\right|\hspace{-0.5mm}\bullet\hspace{-0.5mm}\circ\hspace{-0.5mm}\circ\hspace{-0.5mm}\circ
\end{array}\hspace{-2mm}\right\} $\vspace{1mm}} & {\footnotesize $2$} & {\footnotesize $1$} & {\footnotesize $3$} & {\footnotesize $-2$} & {\footnotesize $-1$} & {\footnotesize $3$} & \textcolor{black}{\scriptsize $(1:1,2,3)_{+}(1,2,3:1)_{-}$}\tabularnewline
\textcolor{blue}{\large ${\color{blue}{\color{blue}\triangleleft}}$}{\scriptsize $\left\{ \hspace{-2mm}\begin{array}{c}
\circ|\hspace{-0.4mm}\bullet\cdots\bullet\hspace{-0.4mm}\circ\hspace{-0.4mm}\bullet\hspace{-0.4mm}\bullet|\bullet\bullet\circ\\
\circ|\hspace{-0.4mm}\bullet\cdots\bullet\hspace{-0.4mm}\bullet\hspace{-0.4mm}\circ\hspace{-0.4mm}\circ|\circ\bullet\circ
\end{array}\hspace{-2mm}\right\} $\vspace{1mm}} & {\footnotesize $1$} & {\footnotesize $0$} & {\footnotesize $4$} & {\footnotesize $-1$} & {\footnotesize $-1$} & {\footnotesize $3$} & \textcolor{black}{\scriptsize $(3:1,2)_{+}(1,2:2)_{-}$}\tabularnewline
\end{tabular}

\caption{Schematic representations of the excited states considered in Fig. \ref{fig6}(b).  
We also present the excitations in terms of $I_{\sigma}$ and $Q_{\sigma}$ [see
Eq. (\ref{eq:exctfinite})]
as well as the notation of Ref. \onlinecite{chicoentropyext2}.}
\par\end{center}
\end{table}

In Fig. \ref{fig6}(b), we show $\Delta S_{1}$ as function of 
$\frac{2\pi^{2}}{3}(\ell/2L)^{2}$
for several excitations, in the regime that $(\ell/L)\ll1$ for a
system with $2L$ sites. The schematic representation of these non-compact
excitations are also presented in terms of $I_\sigma,Q_\sigma$ and $j_\sigma$
in Table II.
The solid lines in Fig.  \ref{fig6}(b) correspond to Eq. (\ref{eq:detas1})
with the values of $d^{\beta}$ shown in parentheses. 
As observed in this figure, the values of $d^{\beta}$ that fit our data to Eq. (\ref{eq:detas1})
are consistent with
$d^\beta=\sum_{\sigma=\pm}(j_\sigma+j_\sigma')$. 
Slightly different than
expected from the Eq. (\ref{eq:detas1}). 
In order to better understand this discrepancy, we also
reproduce the results of the same excited states considered in the Refs. 
\onlinecite{chicoentropy} and \onlinecite{chicoentropyext2} for the XX chain. 
Indeed, we also get the same values of  $d^\beta$ obtained
in these references. For example, for the excitation associated with the 
operator $i\partial\phi$, i. e.  $(1:1)$, we got $d^{\beta}=1$.
However, we did not interpret this value as the scaling dimension 
of the {\emph{primary}} operator $i\partial\phi$ (which is $1$), 
due to the momentum of the excitation $(1:1)$. From the CFT point of view 
this primary operator  has momentum 0. 
However, the excitation $(1:1)$ has momentum $\Delta P=2\pi/L$.
These results show that a deeper analysis of the Rényi entropy of excited
states associated with the descendent operators 
should be done in order to understand the exact dependence of $\Delta S_{1}$,
in the regime that $(\ell/L)\ll1$.
Below, we comment on two points that may help infer the exact form of the factor
$d^\beta$ that appears in  (\ref{eq:detas1}).
First, note that Eq. (\ref{eq:detas1})  was obtained exactly for the primary operators. 
For instance, the operator $i\partial\phi$
is primary in terms of the Virasoro algebra. However, for a
 larger algebra as the Kac-Moody algebra 
(which is the correct one for the XXZ chain) the operator $i\partial\phi$
  can be seen as a descendent operator. So, the result of Eq. (\ref{eq:detas1}) 
  independent  whether the operator $i\partial\phi$ is primary or not. 
Second, our results suggest that the operators associated with configurations 
as the ones depicted in Table II may be express
by a product of vertex operators with scaling dimensions $d_{I,Q}$
times a product of derivatives of the fields $\phi$ and $\bar{\phi}$
with scaling dimension $j+j'$,
 and this would lead a factor $d^{\beta}=j+j'$
  in Eq.  (\ref{eq:detas1}). \cite{privite} 
 Further investigations are needed in order to clarify the exact form of $\Delta S_{1}$
 in the limit $(\ell/L)\rightarrow0$ for a general excited state.

\section{Conclusions}

In this work, we considered a toy model with three spin-interactions
defined on a two-leg geometry which is exactly solvable by means of
a Jordan-Wigner transformation. We were able to obtain analytically
the finite-size corrections of the low-lying energies, and for this
reason we get the exact values of the central charge and the scaling
dimensions. The toy model investigated {[}Eq. (\ref{Eq:Hamiltonian}){]}
is in the same universality class of critical behavior of the $XX$
chain, with central charge $c=1$. Moreover, the scaling dimensions
of the toy model present a similar structure as
the one of the $XX$ chain {[}see Eq. (\ref{eq:scaldim}){]}. 

We also acquired the central charge by analyzing the finite-size corrections
of the Rényi entropy. In the region of the phase diagram where there
is one gapless mode, we get $c=1$. Whereas in the region where there
are two gapless modes, we found that Rényi entropy behaves asymptotically
as
\begin{equation}
S_{\alpha}(L,\ell)=\frac{c_{eff}}{6}\left(1+\frac{1}{\alpha}\right)
\ln\left[\frac{L}{\pi}\sin\left(\frac{\pi\ell}{L}\right)\right]+a_{\alpha},
\end{equation}
where  $c_{eff}=2c=2$. For systems with $n_{gl}$ gapless modes it
is expected that $c_{eff}=n_{gl}c$, at least for non-interaction
systems (see also Fagotti in Ref. \onlinecite{3bodyfagotti} for a
similar discussion). It is interesting to mention that an extension
of the present model for $N$-leg ladders with ($N+1)$ spin interactions
can also be done and would lead to an effective central charge $c_{eff}=N$
(at least for some regions of the parameter space). 
This result shows that the  von Neumann entropy of the $N$-leg spin ladders 
of size $L$ and $L$ gapless modes should behave as $S_{1}\sim L\ln(L)+aL$, this simply argument
shows that entropic area law may be violated in two-dimensional critical
systems. Indeed, violations of the entropic area law was observed
in systems in dimension higher than one \citep{PhysRevB.74.073103,rmp} 

We also study the Rényi entropy of the excited states. Our results
of the Rényi entropy associated with the compacts as well as the non-compact
excitations are in agreement with the recent prediction done by the
authors of Refs. \onlinecite{chicoentropy} and \onlinecite{chicoentropyext2}. 
\begin{acknowledgments}
The authors thank F. C. Alcaraz and G. Sierra for discussions and a careful
reading of the manuscript. The authors also thank J. Ricardo de Mendon\c{c}a
for  carefully reading our manuscript.
This research was supported by the Brazilian
agencies FAPEMIG and CNPq. 
\end{acknowledgments}
\bibliographystyle{apsrev4-1}
\addcontentsline{toc}{section}{\refname}

\begin{thebibliography}{56}
\bibitem {comment2012}%
  Note that in
  the region III, we would naively expect that $\Delta$ would be negative for some
  values of the  coupling constants. However, this is not happen. The sets of
  couplings such that $\Delta<0$, do not belong to the region III.
\end{thebibliography}
%
\end{document}